\documentclass[fleqn,usenatbib]{mnras}

\usepackage{newtxtext,newtxmath}
\usepackage[T1]{fontenc}
\usepackage{ae,aecompl}
\usepackage{graphicx}   % Including figure files
\usepackage{amsmath}    % Advanced maths commands
\usepackage{amssymb}    % Extra maths symbols
\newcommand{\source}{\hbox{PKS J1421-0643}}
\newcommand{\chandra}{\textit{Chandra}}
\newcommand{\hst}{\textit{HST}}
\newcommand{\vla}{\textit{VLA}}
\newcommand{\fermi}{\textit{Fermi}}

\title[The high-redshift quasar \source]{Inverse-Compton scattering in the resolved jet of the
  high-redshift quasar \source}

\author[D.M.~Worrall et al.]
 {D.M.~Worrall$^{1}$, 
M.~Birkinshaw$^{1}$, 
H.L.~Marshall$^{2}$,
D.A.~Schwartz$^{3}$,
A.~Siemiginowska$^{3}$
\newauthor
and J.F.C.~Wardle$^{4}$
\\
$^1$HH Wills Physics Laboratory, University of Bristol, Tyndall
Avenue, Bristol BS8~1TL, UK \\
 $^2$Kavli Institute for Astrophysics and Space Research, MIT, 77 Massachusetts Avenue, Cambridge, MA 02139, USA \\
 $^3$Harvard-Smithsonian Center for Astrophysics, 60 Garden Street, Cambridge, MA 02138, USA \\
 $^4$Department of Physics, Brandeis University, 415 South Street, Waltham, MA 02454, USA 
}

\begin{document}

\label{firstpage}

\maketitle

\begin{abstract}
 Despite the fact that kpc-scale inverse-Compton (iC) scattering of
 cosmic microwave background (CMB) photons into the X-ray band is
 mandated, proof of detection in resolved quasar jets is often
 insecure.  High redshift provides favourable conditions due to the
 increased energy density of the CMB, and it allows constraints to be
 placed on the radio synchrotron-emitting electron component at high
 energies that are otherwise inaccessible.  We present new X-ray,
 optical and radio results from \chandra, \hst\ and the \vla\ for the
 core and resolved jet in the $z=3.69$ quasar \source.  The X-ray jet
 extends for about $4.5''$ (32 kpc projected length).  The jet's radio
 spectrum is abnormally steep and consistent with electrons being
 accelerated to a maximum Lorentz factor of about 5000.  Results argue
 in favour of the detection of inverse-Compton X-rays for modest
 magnetic field strength of a few nT, Doppler factor of about 4, and
 viewing angle of about $15^\circ$, and predict the jet to be largely
 invisible in most other spectral bands including the far- and
 mid-infrared and high-energy gamma-ray.  The jet
   power is estimated to be about $3 \times 10^{46}$ erg
   s$^{-1}$ which is of order a tenth of the quasar bolometric power, for an 
   electron--positron jet.  The jet radiative power is only about
   0.07 per cent of the jet power, with a smaller radiated power ratio if
   the jet contains heavy particles, so most of the jet power is available for heating
   the intergalactic medium.

\end{abstract}


\section{Introduction}
\label{sec:intro}%1

A key finding with \chandra, only possible due to its spatial fidelity
and sensitivity, has been the detection of kpc-scale X-ray jets in
quasars \citep{tananbaum}.  Following the discovery of a jet during
the initial \chandra\ focussing operations \citep{schwartz0637},
several quasar jet surveys have been conducted, normally with
individual exposures in the range 5 to 20 ks, targeting quasars mostly
at low redshift, $z \lesssim 1$, with jets well mapped in the radio
\citep{sambruna02, sambruna04, sambruna06, marshall05, marshall11,
  marshall18, jorstad, hogan11}.  While all the quasar cores are X-ray
detected, the X-ray detection rate for at least
  one resolved jet knot is also high, at roughly 60 per cent.  Some
\chandra-discovered jets have had follow-up observations to study
X-ray morphology and spectral properties, the deepest exposures ($>
100$~ks) being for PKS 1127-145 and 4C+19.44 \citep{siemiginowska1127,
  harris4c19.44}.

The X-ray emission mechanism of these low-redshift quasar jets is
surprisingly hard to ascertain, and remains a subject of heated
debate, having profound implications for the kinetic power of the jets
or the acceleration process, or both \citep[see][]{wrev}.  As in
nearby 3C\,273, at $z=0.158$, whose kpc X-ray jet was known before
\chandra\ \citep{willingale, roser}, there are good arguments for
believing synchrotron emission dominates in high-contrast knots,
requiring those regions to be efficient particle accelerators to
extreme energies \citep[e.g.,][]{jester06, marchenko}.  On the other
hand, the high detection rate in the X-ray of radio-jetted quasars,
often with optical detections or upper limits falling below spectral
interpolations between radio and X-ray, would require an extraordinary
jet acceleration process if from electron
synchrotron radiation \citep[e.g.,][]{schwartz0637,schwartz-fourjets,
  sambruna04}, and has given weight to an interpretation of inverse
Compton (iC) scattering on the cosmic microwave background (CMB)
\citep[i.e., the iC-CMB mechanism,][]{tavecchio00, celotti}. Early
applications of the iC-CMB model stressed that the quasar jet must
remain highly relativistic on kpc scales, such that it sees boosted
CMB in its rest frame and emits beamed X-rays in the observer's frame.
 Modelling has tended to find beaming parameters
  that seem quite extreme for large distances from the core (Doppler
  factors in the range 5 to 20, implying high Lorentz factors and
  small angles to the line of sight), with no significant jet
  deceleration between parsec and kpc scales, sometimes out to many
  hundreds of kpc before subsequent jet bending
\citep[e.g.,][]{jorstad04, tavecchio04}.  The implications are high
jet power, often rivalling the quasar radiative power
\citep{schwartz-fourjets}.

Whereas X-ray synchrotron radiation requires there to be electrons of
extreme (super-TeV) energy, the existence of electrons with low
relativistic energies is certain, as higher-energy electrons lose
energy by radio synchrotron radiation, and so iC-CMB is mandatory.
The increased energy density of the CMB at high redshift should lead
to proportionally more X-ray jet detections at high $z$
\citep{schwartzz}, but statistical tests that have looked for the
expected $(1+z)^4$ dependency in the X-ray to radio flux ratio have
failed to support an exclusive iC-CMB description
without invoking additional evolutionary effects in the jets for which
observational support is unproven \citep{marshall11, marshall18}.  In
a similar vein, it has been argued that if resolved quasar X-ray jets
at $z < 1$ are scaled with the CMB, the addition to the core emission
violates observations in predicting excess quasar X-ray luminosity
evolution \citep{miller11} and much brighter X-ray fluxes
 than are found in targeted SDSS quasars at $z >
4$ \citep{zhu19}.

Clever approaches have been employed to test the iC-CMB model at low
redshift, using \hst\ optical polarization \citep{cara13}, or
\fermi\ upper limits to constrain Compton scattering to high-energy gamma rays
\citep{meyerg, meyer17, breiding}, with both disfavouring the iC-CMB
model in several sources, of which some were
  earlier modelled as iC-CMB.  It seems convincing that it is incorrect to assume
that the resolved jets in low-redshift quasar jets are dominated by
iC-CMB, and incorrect to use all of the measured
  X-ray flux to scale them to high redshift. A more promising
approach is to look in depth directly at jets at high redshift, with a
view to inferring their likely physical conditions and beaming
parameters.  Scaling to low redshift might then
  provide an estimate of the contribution of iC-CMB to the X-ray
  emission of low-redshift quasars.

A lack of jets with strong X-ray and weak radio emission has been seen
as a problem for the iC-CMB model.  How large a problem remains
unclear, because surveys targeted to find resolved X-ray jets are
biased towards quasars with strong radio jets.  In a 19-ks
\chandra\ exposure, \citet{simionescu0727} found serendipitously an
extended X-ray jet in the $z=2.5$ quasar B3\,0727+409, whose radio
emission is point-like except for one knot $1.3''$ from the core.
They conclude strong support for the iC-CMB mechanism.  The space
density of radio-loud quasars at moderate to high redshift that have
high-fidelity radio observations, coupled with the sky coverage of
\chandra, is sufficiently low that it is not unexpected that we have
failed to find additional such cases in searches of \chandra\ archival
data. More recent \chandra\ targetted observations
  of 14 previously unobserved quasars have found two candidate X-ray
  jet systems where no underlying radio jet emission is seen
\citep{schwartz-an}.

\begin{figure}
\centering
\includegraphics[width=0.9\columnwidth]{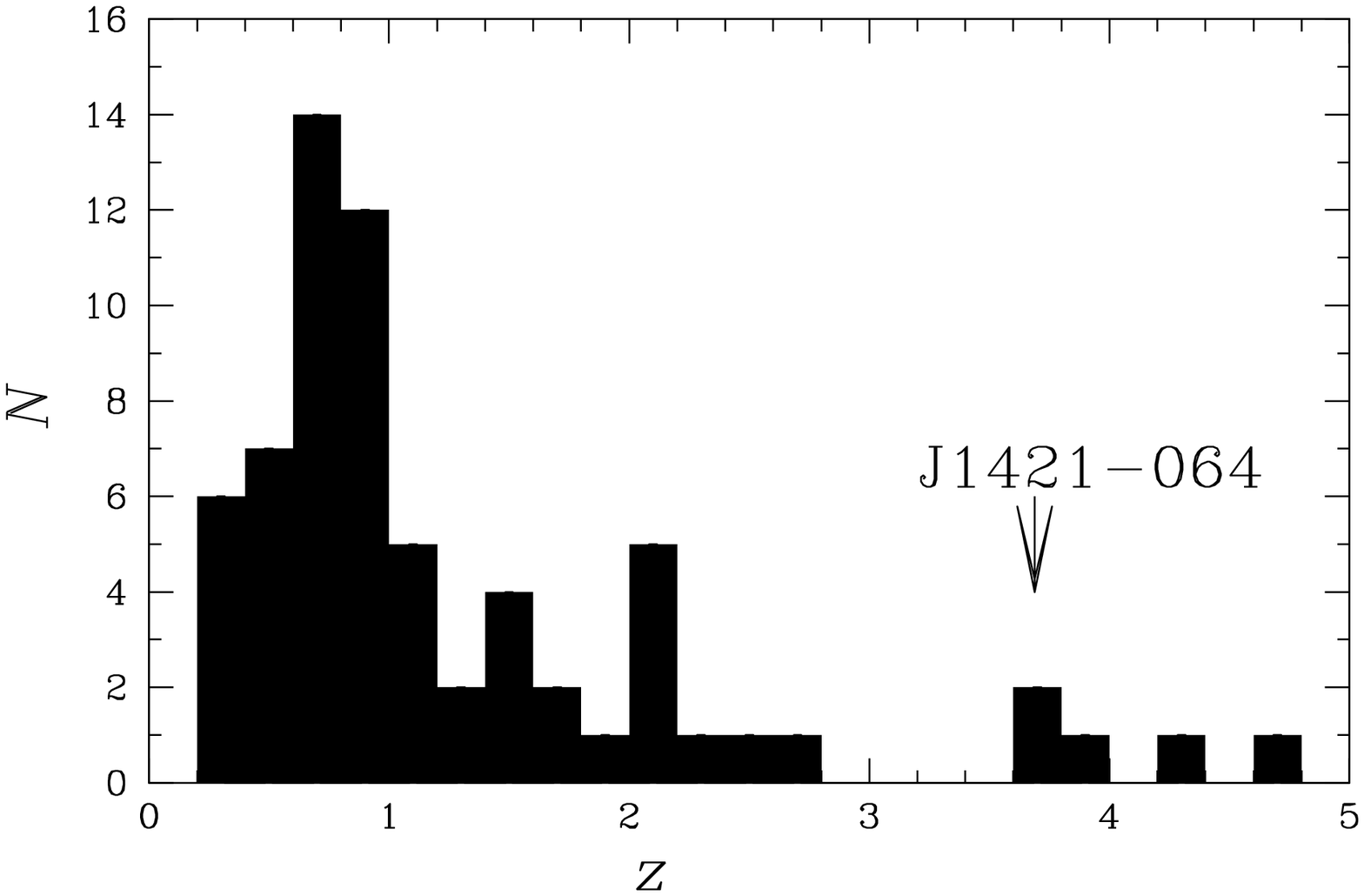}
\caption{ Redshift distribution of 66 quasars at $z > 0.2$ with
  detected kpc-scale X-ray jets.  Objects here are in one
  or more of the surveys reported by \citet{sambruna02,  sambruna04,
    marshall05, marshall11, marshall18, jorstad, hogan11,
mckeough},
% (although not necessarily first reported in
%  those papers), 
and eleven additional cases have been added from
\citet*{schwartz0637, chartas02, fabian0020, siemiginowska0738,
  siemiginowska1127, cheung1745, millerPG1004, tavecchio1229, 
   miller4c65, saez, simionescu0727}.  The median redshift is 0.89.
}
\label{fig:histo}
\end{figure}

That most jet work to date has concentrated on nearby objects is seen
from Figure~\ref{fig:histo} which displays a histogram (shaded) of the
redshift distribution of all quasars at $z > 0.2$ for which X-ray
kpc-scale jets are reported to our knowledge. Because observations
must be with \chandra\ and, with the exception of B3\,0727+409, were
all as targets, there remain relatively few detections, particularly
at medium to high redshift. However, detection rates in short
exposures at the higher redshifts are high.  For example,
\citet{mckeough} summarize \chandra\ results for a sample of eleven
quasars at $z > 2.1$ with resolved radio jets, five of which are at $z
> 3.5$.  Eight of the eleven display extended X-ray emission
associated with jet features, and in seven of these cases the
\chandra\ exposure times were short, between 3.3 and 20.1 ks.  Three
of the $z > 3.5$ sources show jet features, as do additional sources
at $z=3.89$ and $z=3.63$ reported by
  \citet{cheung1745} and \citet{saez}, respectively, giving the five
  known jet sources at $z > 3.5$ listed in Table~\ref{tab:highz}.  Of
  these five, the source with the longest X-ray jet but shortest
observation time is \source.  We have therefore undertaken a programme
of deeper observations of \source, using \chandra, \hst, and the \vla,
with a view to testing the plausibility and implications of the iC-CMB
process to explain its X-ray emission, and measuring the physical
parameters of this high-redshift jet.

%------------ begin figure
%------------ end figure

%----------- begin table
\begin{table}
\caption{Archival X-ray quasar jets at $z > 3.5$.}
\label{tab:highz}
\setlength\tabcolsep{2.0pt}
\begin{tabular}{llcl}
\hline
(1) & (2) & (3) & (4)\\
Name
  & $z$ 
  & \chandra\ 
  & References \\
  & 
  & (ks)
  &  \\
\hline
 PMN J2219-279 & 3.63 & 46.0 & \citet{saez}\\
\source\  & 3.689  & 3.3  & \citet{cheung1421}\\
          &        &      & \citet{mckeough}\\
4C+62.29  & 3.89  & 18.3  & \citet{cheung1745}\\
GB 1508+5714  & 4.3  & 89.0 & \citet{yuan1508} \\  
          &        &      & \citet{siemiginowska1508}\\
          &        &      & \citet{mckeough}\\
GB 1428+4217 & 4.72  & 10.6     & \citet{cheung1428}\\
          &        &      & \citet{mckeough}\\
\hline
\end{tabular}
\begin{minipage}{\linewidth}
(3) Reported \chandra\ exposure time.
\end{minipage}
\medskip
\end{table}
%----------- end table

In this paper we adopt values for the cosmological parameters of $H_0
= 70$~km s$^{-1}$ Mpc$^{-1}$, $\Omega_{\rm {m0}} = 0.3$, and
$\Omega_{\Lambda 0} = 0.7$.  Thus at \source's redshift of $z =3.689$
\citep{hook}, 1~arcsec corresponds to a projected distance of
7.18~kpc. Spectral index, $\alpha$, is defined in the sense that flux
density is proportional to $\nu^{-\alpha}$.  In the X-ray,
$\alpha$ relates to the modelled photon
spectral index, $\Gamma$, as $\Gamma = \alpha + 1$.

\section{Observations and reduction methods}
\label{sec:obs}%2

\subsection{\chandra\ X-ray}
\label{sec:xobs}

Our observations of \source\ with the S3 back-illuminated CCD of the
Advanced CCD Imaging Spectrometer (ACIS) on board \chandra\ were made
as given in Table~\ref{tab:chandraobs} (ObsIDs 20445, 21054, 21055,
21056 and 21057).  In addition we make use of archival data from a
short observation taken with the same CCD eleven years earlier (ObsID
7873), also listed in Table~\ref{tab:chandraobs}.  Details of the ACIS
instrument and its modes of operation are given in the
\chandra\ Proposers' Observatory Guide\footnote{
  http://cxc.harvard.edu/proposer}.  Results presented here use {\sc
  ciao} \citep{fruscione} {\sc v4.11} and the {\sc caldb v4.8.2} calibration database.  We
reprocessed the data following the software `threads' from the
\chandra\ X-ray Center (CXC)\footnote{ http://cxc.harvard.edu/ciao},
to make new level~2 events files, applying the sub-pixel event
repositioning algorithm. VFAINT screening was not applied for the
analysis here due to the small angular scale of the quasar and its
jet, and the fact that such screening can remove valid events from the
core.

%----------- begin table
\begin{table}
\caption{\chandra\ Observations of \source.}
\label{tab:chandraobs}
\begin{tabular}{rlccr}
\hline
(1) & (2) & (3) & (4) & (5)\\
ObsID
  & Date 
  & Frame
  & Sub-
  & Exposure \\
  & 
  & Mode
  & array
  & (ks) \\
\hline
 20445  & 2018 Mar 27  & VFAINT & 1/4 & 22.001\\
 21054  & 2018 Mar 29  & VFAINT & 1/4 & 41.033 \\
 21055 & 2018 Mar 30 & VFAINT & 1/4 & 24.488 \\
 21056  & 2018 Mar 31  & VFAINT & 1/4 & 18.559 \\
 21057  & 2018 Apr 01  & VFAINT & 1/4 & 11.766 \\
 7873  & 2007 Jun 04  & FAINT & 1/8 & 3.342 \\
\hline
\end{tabular}
%\begin{minipage}{\linewidth}
%(4) \chandra\ exposure time after flare screening.
%\end{minipage}
\medskip
\end{table}
%----------- end table

The observations were free from large background flares and, after
removal of time intervals when the background deviates more than
$3\sigma$ from the average value, the total exposure time was 121.2
ks, distributed as given in Table~\ref{tab:chandraobs}.  We applied
small corrections to align the centroid of 1--5-keV counts from the
core of \source\ to our measured radio core position of RA$=14^{\rm h}
21^{\rm m} 07^{\rm s}\llap{.}756$, Dec$=-06^\circ 43' 56''\llap{.}36$.
No observation required a shift larger than 0.17 arcsec.

Our spectral fitting used {\sc
  xspec}\footnote{https://heasarc.gsfc.nasa.gov/xanadu/xspec/}
v12.10.1d.  If there were more than 300 net counts in the region of
interest we used the $\chi^2$ statistic after grouping data to a
minimum of 20 counts bin$^{-1}$; otherwise we used cstat,
after binning to at least one count per bin, and
  where the wstat statistic gives a rough indication of goodness of
  fit.  For absorption we used the phabs and zphabs model for
Galactic and intrinsic absorption, respectively, with the abundance
table of \citet{asplund}.  \source\ is at a Galactic latitude of
$50^\circ$, and the line-of-sight column density is given as $N_{\rm
  H} = 3.4 \times 10^{20}$ cm$^{-2}$ by the CXC {\sc colden} routine
using data of \citet{dlock90}.  This component of fixed Galactic
absorption was included in all spectral modelling. Local background
was used with exclusion regions corresponding to sources found by the
{\sc ciao wavdetect} task. Extracted spectra and response files were
combined using the {\sc ciao combine\_spectra} task and fitted to
spectral models over the 0.5 to 7-keV energy band.  The ACIS response
was significantly higher at low energies when the earlier short
observation was made, but we find negligible change to fitted spectral
shape whether or not these earlier data are included.  The X-ray
spectrum of the core was measured from a circle of radius 1.25 arcsec
using the {\sc ciao specextract} task, correcting for missing counts
based on the Point Spread Function (PSF) and sampling background from
a source-centred annulus of radii 10 and 50 arcsec.  The jet spectrum
was measured using regions described in Section \ref{sec:jetspec}.

After the X-ray spectrum of the core had been modelled, we used the
CXC {\sc SAOsac raytrace} and {\sc marx} software to make 50
simulations of the PSF using inputs appropriate for the longest
observation, ObsID 21054, including supplying the
  aspect solution and applying the EDSER subpixel algorithm to match
  the data.  These were combined to provide an image of the spatial
extent of the PSF which was convolved with models for spatial
comparison with data.

\subsection{\vla\ radio}
\label{sec:robs}

We observed \source\ in programme SJ0472 for 3.5~h with the Karl
G.~Jansky Very Large Array (\vla) in its A-array configuration on
2018~March~04. Data were taken in the C and X bands, covering 3.98 to
12.02 GHz. Flux-density calibration was based on 3C\,286 (J1331+3030)
which also provided the polarization position angle reference. The
low-polarization calibrator OQ\,208 (J1407+2827) provided measurements
of polarization leakage. Phase calibration used J1408-0752, which lies
3.3$^\circ$ from \source. Calibration and mapping were performed
within the {\sc CASA} software.

\begin{figure}
\centering
\includegraphics[width=1.0\columnwidth]{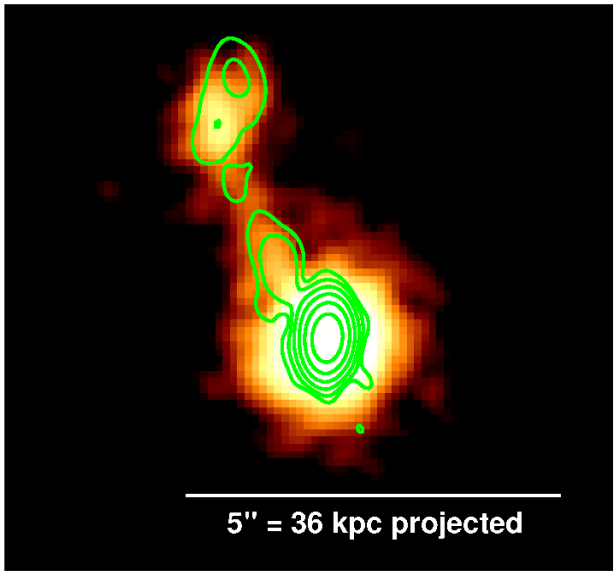}
\caption{ 0.5--5-keV\chandra\ image of \source\ in 0.0984-arcsec pixels,
  smoothed with a Gaussian of $\sigma$=1.5 pixels.
Radio contours are from a 5.13-GHz map made using data from
\vla\ programme SJ0472 with a restoring beam of 0.48 arcsec $\times$ 0.28
arcsec, increasing by factors of 4 from 60 $\mu$Jy beam$^{-1}$.
}
\label{fig:xr5}
\end{figure}

Extensive flagging for interference was required, with about 30 per
cent of the data rejected to achieve good calibration and reliable
maps. Mapping was further complicated by the presence of a bright
source near the edge of the primary beam of the \vla\ antennas in the
C~band, but outside the primary beam at X~band, and by inaccuracies in
the bandpass calibration. Independent maps at lower and higher
frequencies, and with suitably-chosen groups of reliable channels,
resulted in excellent final maps centred at 5.13, 7.00, 9.00, and
10.7~GHz which were used independently, or combined, for spectral and
structural analyses. Contours of the 5.13-GHz map, which has a noise
of 9.1 $\mu$Jy beam$^{-1}$ with a synthesized beam of 0.48 arcsec
$\times$ 0.28 arcsec, are shown in Figure~\ref{fig:xr5}.

%------------ begin figure
%------------ end figure

We also made maps from the 1.4 and 5-GHz data for \source\ from
archival \vla\ dataset AC755 \citep{cheung1421}, taken on 2004 December 22. The \vla\ was
in the A-array configuration for these observations for which about 26
and 10 min of data were taken at 1.4 and 5 GHz, respectively.  Similar
angular resolution (at 5~GHz) but poorer sensitivities were achieved
from these earlier observations which also used 3C\,286 as the primary
flux-density calibrator.

\subsection{\hst}
\label{sec:hstobs}%

We observed \source\ with the WFC3/UVIS instrument on \hst\ on 2018
February 22.  The source was observed using the F555W (pivot
wavelength 530.595 nm) and F814W (pivot wavelength 804.810 nm) filters
(program GO-15376) in three and two separate dithered exposures
totalling 1.35 and 0.98 ks, respectively.  Source orientation was such
that the jet direction extends between quasar diffraction spikes.  We
made final images for each filter using the pipeline-processed files
corrected for charge transfer efficiency as input to the {\sc python
  drizzlepac tweakreg} and {\sc astrodrizzle} software routines,
guided by the Space Telescope Science Institute (STScI)
tutorials\footnote{https://spacetelescope.github.io/notebooks/.}.  To
correct for Galactic extinction the F814W and F555W flux densities
have been increased by factors of 1.07 and 1.129, respectively, based
on values from the NASA/IPAC Extragalactic Database (NED).
Flux-density upper limits for undetected jet features are based on the
mean and standard deviation of values found through placing the region
of interest at multiple locations across the CCD, and are quoted at
$3\sigma$ significance.

\section{Jet results}
\label{sec:results}%3

\subsection{Imaging}
\label{sec:imaging}

\begin{figure}
\centering
\includegraphics[width=0.85\columnwidth]{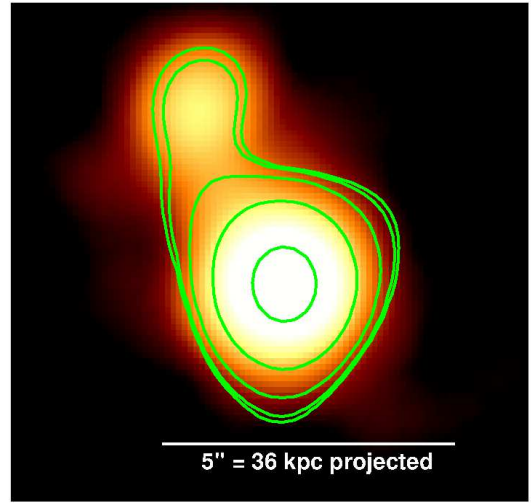}
\caption{ 0.5--5-keV\chandra\ image of \source\ in 0.0984-arcsec pixels,
  smoothed with a Gaussian of $\sigma$=5 pixels.
Radio contours are from the 1.42-GHz map made using data from
\vla\ programme AC755I with a restoring beam of 1.54 arcsec $\times$ 1.27
arcsec, logarithmically spaced between 3 and 200 mJy beam$^{-1}$.
}
\label{fig:xr1.4}
\end{figure}

Figure~\ref{fig:xr5} shows that \chandra\ detects a continuous jet
extending NNE of the quasar core.  After a bright knot that
corresponds with 5-GHz radio emission the X-rays fade and the jet
appears to kink slightly before a bright terminal region where the
X-rays are brightest to the SE of a prominent 5-GHz knot.  When
compared with lower-resolution 1.4-GHz data, the X-ray and radio show
good overall correspondence (Fig.~\ref{fig:xr1.4}).

%------------ begin figure
%------------ end figure

\begin{figure}
\centering
\includegraphics[width=0.95\columnwidth]{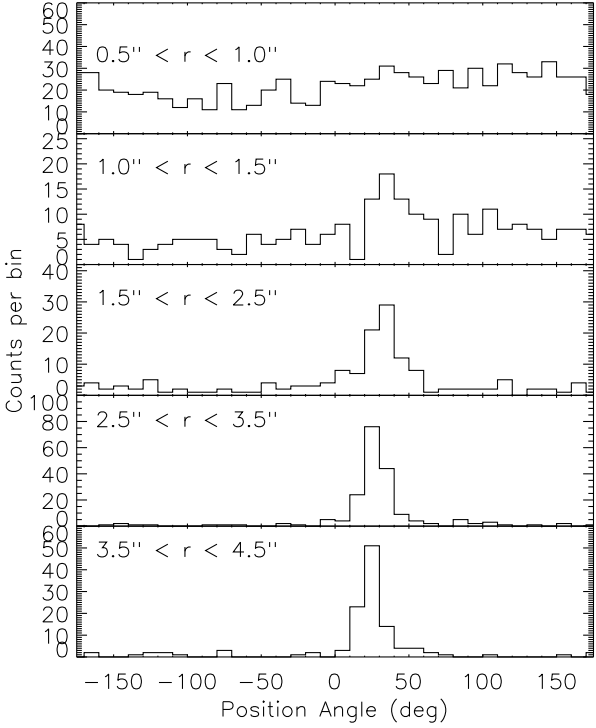}
\caption{ Azimuthal distributions of the source-centred 0.5--5-keV
  X-ray counts in five radial bands.  No background is subtracted.
  Errors should be taken as Poissonian.
}
\label{fig:hm-pa}
\end{figure}

\begin{figure}
\centering
\includegraphics[width=0.95\columnwidth]{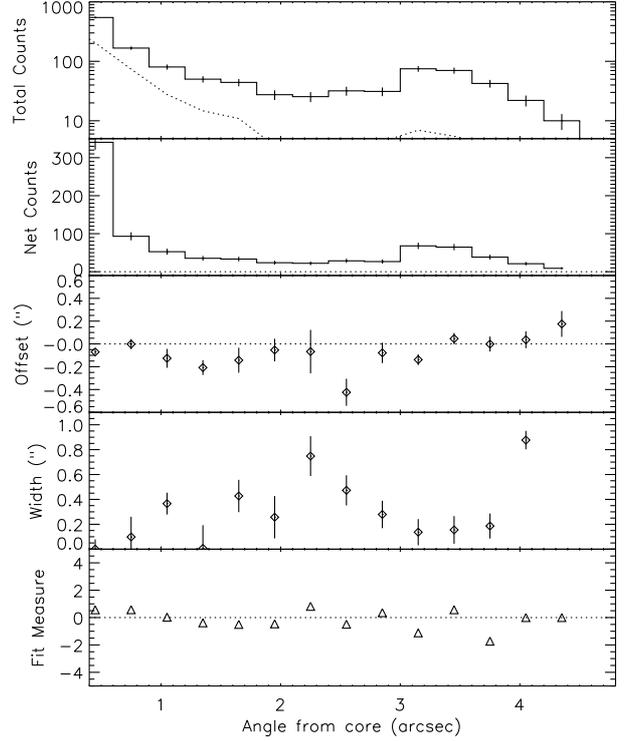}
\caption{ Fitted transverse structural parameters of the 0.5--5-keV
  X-ray jet measured in radial bins of length 0.3 arcsec.  Panels from
  top to bottom show: 1. Total counts (histogram) compared with
  background from mirror-image counterjet direction (dashed line).
  2. Net counts.  3. Offset from reference position angle of 25
  degrees. 4. Excess width, defined as the sigma of the Gaussian
  needed to be added in
  quadrature to that of a point source in order to fit the
  data. 5. Fit measure, based on a Kolmogorov-Smirnov (K-S) test, with
  the K-S probability of achieving a better fit expressed in Gaussian
  standard deviation units, so that negative values correspond to
  better-than-average fits and positive values to worse-than-average
  fits.}
\label{fig:hm-trans}
\end{figure}

For a quantitative investigation of the structure of the X-ray jet,
starting from the reprocessed events file we first made azimuthal
distributions of the counts in five radial bands
(Fig.~\ref{fig:hm-pa}).  The upper panel of this figure shows that
PSF asymmetry may be preventing detection of the X-ray jet
within 1 arcsec of the core.  No excess counts are seen in the CCD
readout direction, roughly orthogonal to the jet direction, proving
that the 1/4-sub-array used for the observations was sufficient to
mitigate any significant pile-up in the core emission.  The lower
panels show the jet position angle shifting from between 30 and 40
degrees to between 20 and 30 degrees with increasing radial distance.
We then placed a broad rectangle over the jet at a position angle of
25 degs (so centred on the jet where it is bright at about 3.7 arcsec
from the core), and the rectangle's mirror-image in the counterjet
direction to sample background. In radial bins of length 0.3-arcsec we
have used a maximum-likelihood method to fit the centroid and width of
the transverse profile of jet counts, with results shown in
Figure~\ref{fig:hm-trans}. The middle panel quantifies the jet
bending.  Importantly, the panel below it finds the fainter parts of
the jet in particular are detectably broadened, although emission at
about 1.35 arcsec is unresolved in the transverse direction.

%------------ begin figure
%------------ end figure

%------------ begin figure
%------------ end figure

\begin{figure*}
\centering
\includegraphics[width=5.0truein]{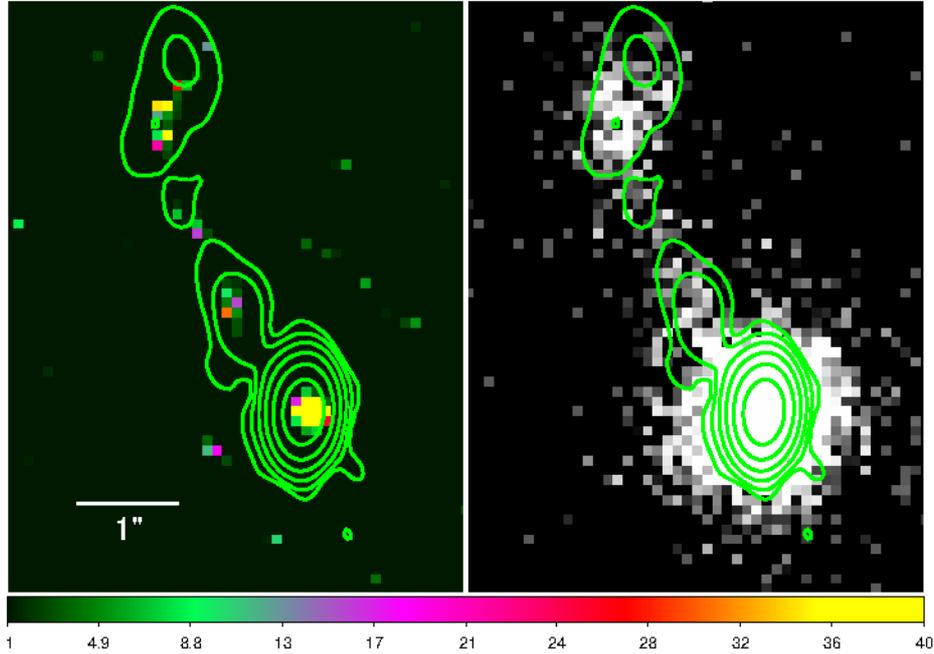}
\caption{0.5--5-keV\chandra\ images in 0.0984-arcsec pixels with
  radio contours from Fig.~\ref{fig:xr5}.
  {\bf Left}: after PSF deconvolution
  showing the core and bent jet.  The
  colour bar is linear in units of counts. The scale is chosen to show
  counts in features in the jet and the core is strongly saturated.
  {\bf Right}: The unsmoothed data for reference.
}
\label{fig:arestore}
\end{figure*}

For further investigation of the structure of the X-ray jet we used
the {\sc ciao arestore} task to deconvolve the data and modelled PSF
using the Richardson Lucy deconvolution algorithm.  The provided
documentation warns that the algorithm has a tendency to break faint
continuous features into points, but it is nonetheless useful in our
case (see Fig.~\ref{fig:arestore}) in confirming the direction of
bending.  The emission at 1.35 arcsec from the core, found in
Figure~\ref{fig:hm-trans} to be unresolved in the transverse
direction, is confirmed from the analysis here to be dominated by a
distinct knot at a position angle of $37^\circ$ that is resolved along
the jet but less resolved in the transverse direction.

%------------ begin figure
%------------ end figure

\begin{figure}
\centering
\includegraphics[width=1.0\columnwidth]{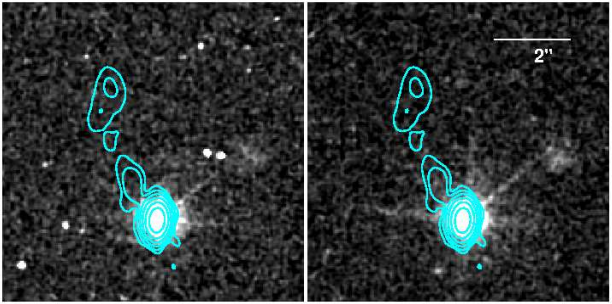}
\caption{ HST images (left F814W filter, right F555W filter) with
  radio contours from Fig.~\ref{fig:xr5}. 
The pixel size is 0.0395 arcsec and the images are
  smoothed with a Gaussian of $\sigma$=2 pixel.
A diffuse source unrelated to the quasar is seen in both images about
3 arcsec to its NW, below the extension of the linear charge-transfer artefact that
lies between the diffraction spikes.}
\label{fig:hst}
\end{figure}

Unlike the radio and X-ray, clear jet emission is not apparent in the
optical  (Fig.~\ref{fig:hst}).  From visual inspection, the most
prominent feature close to the quasar is a diffuse source seen in the
data at both
frequencies about 3 arcsec to its NW. 

%------------ begin figure
%------------ end figure

\subsection{Spectra}
\label{sec:jetspec}

\begin{figure}
\centering
\includegraphics[width=0.85\columnwidth]{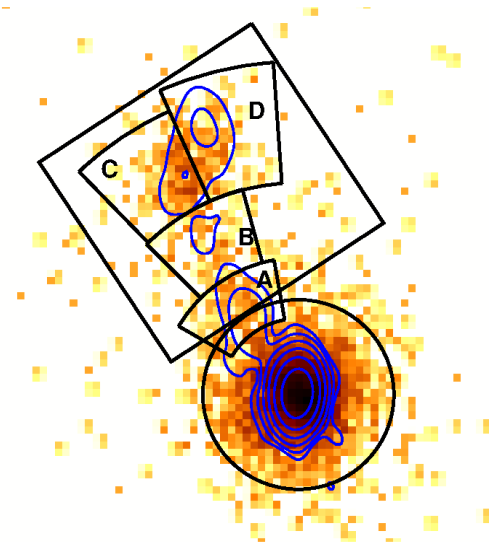}
\caption{ Unsmoothed X-ray data with radio contours of Fig.~\ref{fig:xr5},
showing regions used for data extraction.}
\label{fig:jetregions}
\end{figure}

\begin{figure}
\centering
\includegraphics[width=0.55\columnwidth]{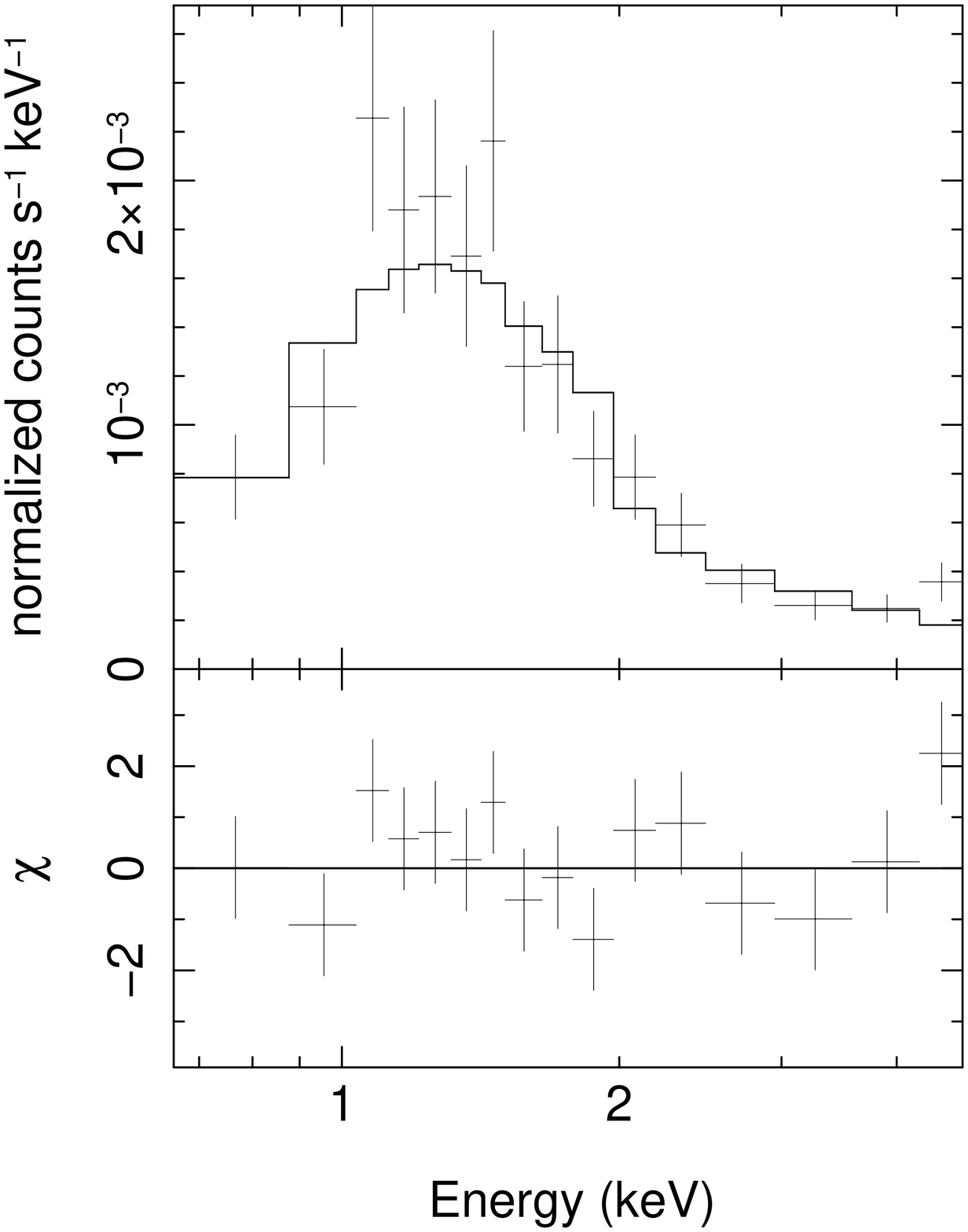}
\includegraphics[width=0.40\columnwidth]{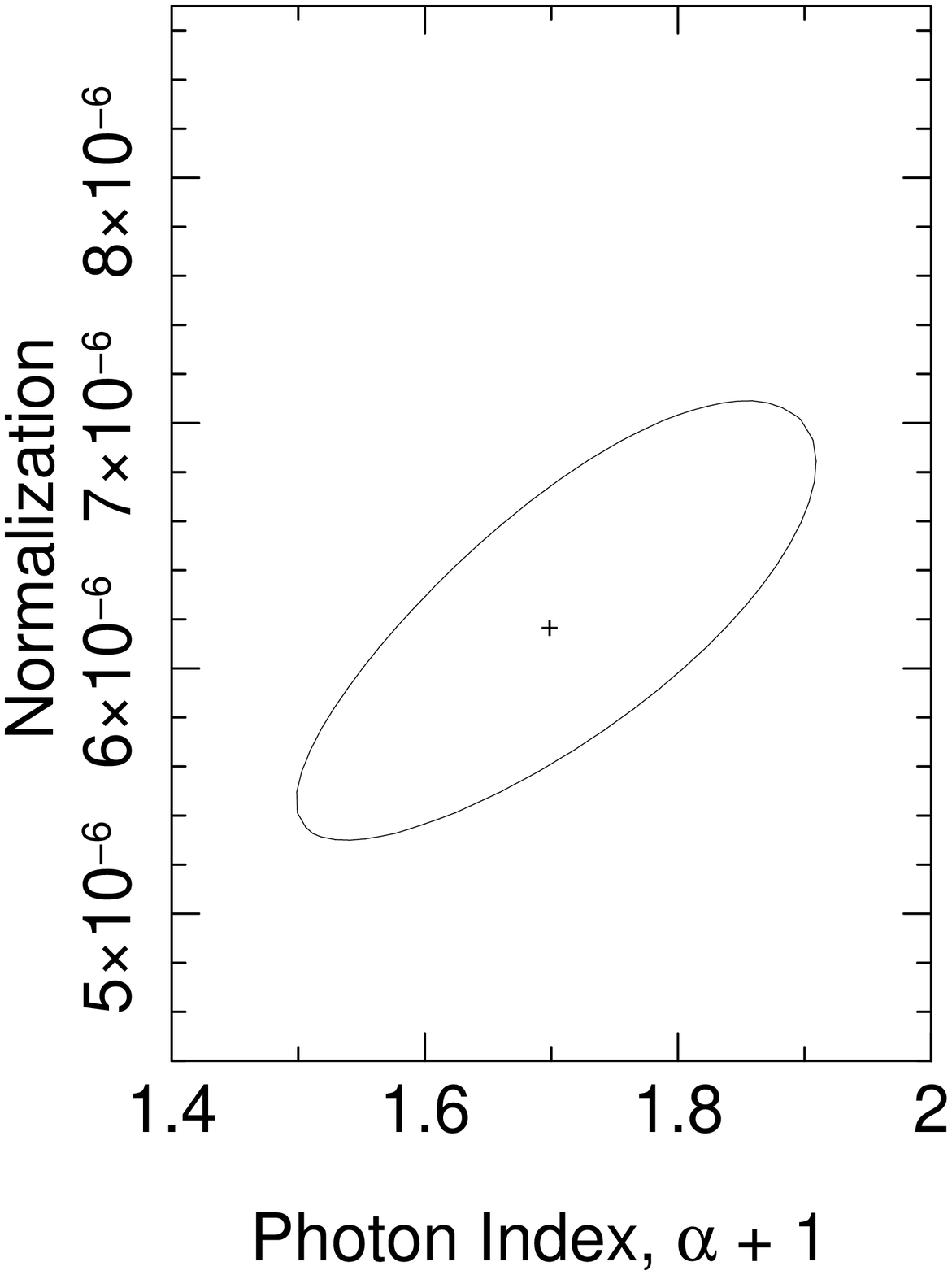}
\caption{ Left: X-ray spectrum of the jet fitted to a power law with
  no absorption in excess of the Galactic value (Table~\ref{tab:xspec}).
Right: Uncertainty contour in spectral index and 1~keV-normalization
(photons cm$^{-2}$ s$^{-1}$ keV$^{-1}$) shown at the
level of 90\% confidence for one interesting parameter.}
\label{fig:jetspec}
\end{figure}

We have fitted X-ray spectra extracted from several regions shown in
Figure~\ref{fig:jetregions}. The large rectangular region samples the
overall jet spectrum, excluding emission from the core, and
chi-squared fitting finds those data fit well a single-component power
law with no absorption in excess of Galactic: $\chi^2 = 16.3$ for 14
degrees of freedom (dof).  The spectrum is shown in
Figure~\ref{fig:jetspec}.  Assuming isotropic radiation, the jet's
source-frame luminosity, corrected for Galactic absorption, is $3.2
\times 10^{45}$ erg s$^{-1}$ (2--20) keV, but see
  the relativistic beaming discussion in Section
  \ref{sec:jetmodellingpower}.

%------------ begin figure
%------------ end figure

%------------ begin figure
%------------ end figure

Our sub-regions have been chosen with reference to the combined
intensity in the X-ray and radio.  They are defined as pie-slice
segments of annular bands, and for computation of volume the symmetry
axis is assumed to be the radius vector from the core that dissects
the pie-slice into two equal parts.  Jet region A incorporates the
first radio and X-ray knot.  In region B, despite a radio knot, the
X-ray surface brightness drops.  The final feature is divided into
region C which includes the brighter X-ray part of the emission, and
region D where a terminal radio feature
  corresponds to a drop in X-ray surface brightness, although the
  boundaries are chosen to give similar numbers of net counts in
  regions C and D.  This means that the impact of the lower
  X-ray-to-radio ratio at the extremity of the jet is not fully
  accommodated in our subsequent modelling.  The annulus is
completed, with jet region excluded, to sample the background for each
extraction, resulting in the best possible exclusion of (weak)
contributions from the wings of the quasar point spread function
(PSF).  We note that there is necessarily some blending of jet
features due to the PSF. Spectral results are given in
Table~\ref{tab:xspec}, where it can be seen that X-ray spectral index
is found to be constant down the jet, within uncertainties.

%----------- begin table
\begin{table*}
\caption{X-ray power-law spectral fits.}
\label{tab:xspec}
\renewcommand{\arraystretch}{1.3}
\begin{tabular}{lrcccc}
\hline
(1) & (2) & (3) & (4) & (5) & (6)\\
Region
  &  Net counts
  & $\alpha_{\rm x}$ 
  & 1-keV normal-
  & $N_{\rm H_{\rm int}}$
  & $\chi^2$/dof \\
  &  (0.5-10 keV)
  &
  & ization (nJy)
  & ($10^{22}$ cm$^{-2})$
  &  \\
\hline
 Core  & 2748 & $0.78\pm0.10$  & $40.5^{+4.8}_{-4.1}$ & $6.7^{+3.4}_{-2.9}$ & 95.4/105\\
 Jet  & 339 & $0.70\pm0.20$  & $4.1\pm0.6$ & --- & 16.3/14 \\
 ...  & ... & $0.65\pm0.19$  & $4.2\pm0.6$ & --- & 18.6/14* \\
 Jet A  & 38 & $0.53\pm0.86$  & $0.46^{+0.33}_{-0.26}$  & --- & 25.5/29* \\
 Jet B  & 51 & $0.27\pm0.50$  & $0.50^{+0.23}_{-0.18}$  & --- & 44.5/47* \\
 Jet C  & 103 & $0.62\pm0.35$  & $1.23^{+0.33}_{-0.28}$  & --- & 83.6/80* \\
 Jet D  & 96 & $0.70\pm0.35$  & $1.22^{+0.34}_{-0.29}$  & --- & 87.7/70* \\
\hline
\end{tabular}
\medskip
\begin{minipage}{\linewidth}
(3), (4) and (5) Uncertainties are 90\% for 1 interesting parameter. 
(6) $\chi^2$ goodness of fit unless *, for which the wstat statistic associated with cstat is given.
\end{minipage}
\end{table*}
%----------- end table

The radio and optical flux densities for the jet regions are given in
Table~\ref{tab:fd}.  Tables~\ref{tab:xspec} and \ref{tab:fd} also
include values for the quasar nucleus which is discussed further in
Section~\ref{sec:core}.  The jet radio flux densities correspond to
unexpectedly steep spectra in all regions, while the core gives a
typical flat spectrum ($\alpha_{\rm r} \approx 0.4$) characteristic of
partial self-absorption.  The radio jet spectra are steeper even than
expected for synchrotron-emitting electrons above the energy-loss
break in a continuous injection model.  Indeed, the spectral shapes,
 which for regions B, C, and D show clear
  steepening with increasing frequency within the measured band
  (Table~\ref{tab:fd}), match emission from within the synchrotron
exponential tail of an electron spectrum with a maximum cut-off
energy, as shown in our modelling below.

A comparison of the archival (2004) 1.4-GHz and our new (2018)
5.13-GHz data gives a spectral index of $1.3\pm0.1$ for that part of
the jet that can be cleanly separated from the core at the resolution
of the 1.4-GHz data.  This shows that the spectrum of the jet
continues to flatten to lower radio frequencies in the manner modelled
in Section~\ref{sec:jetmodellingresults}.

%----------- begin table
\begin{table*}
\caption{Radio and optical flux densities, $S$, and volumes for the
  extraction regions.}
\label{tab:fd}
\setlength\tabcolsep{3.0pt}
\begin{tabular}{lccccccccc}
\hline
(1) & (2) & (3) & (4) & (5) & (6) & (7) & (8) & (9) & (10)\\
Region
  & $S(5.13\rm~GHz)$
  & $S(7.00\rm~GHz)$
  & $S(9.00\rm~GHz)$
  & $S(10.68\rm~GHz)$ 
  & $\alpha_{\rm r_1}$ 
  & $\alpha_{\rm r_2}$ 
  & $S(3.73 \times 10^{14}$
  & $S(5.65 \times 10^{14}$
  & Volume\\
  & (mJy)
  & (mJy)
  & (mJy)
  & (mJy)
  &&
  & Hz) ($\mu$Jy)
  & Hz) ($\mu$Jy)
  & (arcsec$^3$)\\
\hline
 Core   & $237.5\pm 1.4$  & $217.4\pm1.0$ & $193.3\pm0.4$ &  $180.9\pm0.7$ &$0.28\pm 0.02$& $0.44\pm 0.01$ & $134\pm3$  & $82\pm3$ & -- \\
 Jet A  & $1.21\pm0.01$ & $0.71\pm0.01$ & $0.46\pm0.01$ & $0.30\pm0.02$ &$1.72\pm 0.05$& $1.7\pm 0.1$ & $< 0.39$ & $< 0.37$ & 0.95\\
 Jet B  & $0.25\pm0.02$ & $0.11\pm0.01$ & $0.01\pm0.02$ & $0.02\pm0.02$ &$2.6\pm 0.4$& $4.0\pm 2.4$ & $< 0.48$ & $<0.24$ & 1.15\\
 Jet C  & $0.39\pm0.02$ & $0.26\pm0.01$ & $0.07\pm0.02$ & $0.01\pm0.02$ &$1.3\pm 0.2$& $5.2\pm 1.1$ & $< 0.33$ & $<0.13$ & 1.49\\
 Jet D  & $0.92\pm0.02$ & $0.50\pm0.01$ & $0.24\pm0.02$ & $0.08\pm0.02$ &$1.9\pm 0.1$& $2.9\pm 0.3$ & $< 0.67$ & $<0.24$ & 2.01\\
\hline
\end{tabular}
\medskip
\begin{minipage}{\linewidth}
Uncertainties are $1\sigma$.  Upper limits are $3\sigma$.  (6) $\alpha_{\rm r_1}$ is two-point radio spectral index 5.13 to 7 GHz.  (7) $\alpha_{\rm r_2}$ is two-point radio spectral index 7 GHz to either 9 or 10.68 GHz as chosen to give the smaller uncertainty. 
\end{minipage}
\end{table*}
%----------- end table

\section{Jet modelling and discussion}
\label{sec:jetmodelling}%4

\subsection{Radio synchrotron emission and detection of $\gamma_{\rm max}$}
\label{sec:jetmodellingradio}%4

We have modelled the synchrotron emission following
\citet*{hardsynch}, where the electron spectrum extends between
Lorentz factors $\gamma_{\rm min}$, for which we adopt a value of 10,
and $\gamma_{\rm max}$.  The value of $\gamma_{\rm max}$ is
unimportant to the production of X-ray iC radiation, and is normally
taken to be any value higher than that needed to produce the observed
radio synchrotron emission and any emission at higher frequencies that
is consistent with a power-law extension.  However, in \source, a
match to the exponentially falling level of radio emission gives a
measurement of $\gamma_{\rm max}$ that depends only on intrinsic
magnetic field strength, $B$ (see
Section~\ref{sec:jetmodellingresults}).  

The measured value of $B$ depends on
relativistic beaming and how close the radio source is to being at
minimum energy.  After passing through terminal hotspots, jet plasma
no longer has bulk relativistic motion and forms lobes seen at highest
contrast to the jets for sources whose jets are in the plane of the
sky, i.e., radio galaxies.  X-ray lobe iC detections in powerful radio
galaxies find magnetic fields that are within a factor of a few of
minimum energy for no proton contribution
\citep{croston05}, lending support to the same state of play in their
kpc-scale jets.  We therefore adopt minimum energy ($B=B_{\rm me}$),
noting that this has been the common assumption in modelling kpc-scale
quasar emission although sometimes assuming equal
  energy density in protons and electrons
  \citep[e.g.,][]{schwartz-fourjets}, which would increase $B_{\rm
    me}$ by the small factor of $2^{(1+\alpha)/3}$ for synchrotron
  emission of spectral index $\alpha$ (in our case $\alpha=0.65$, see
  below).  An alternative common choice is to use equipartition
  magnetic field strengths, $B_{\rm eq}$, and we note that
\begin{equation}
B_{\rm eq} = B_{\rm me} \sqrt{(3+\alpha)/2(1+\alpha)} .
\label{eq:Bmintoeq}
\end{equation}

\subsection{Modelling rationale}
\label{sec:jetmodellingrationale}%4

In modelling \source\ we are strongly influenced by
its high redshift.  The energy density of the
CMB, which increases as $(1 + z)^4$, is
thus likely to dominate that of the magnetic field.  Written in SI
units\footnote{1 nT = 10 $\mu$G}, the CMB energy density is the larger of the two if
\begin{equation}
B < 0.32 ~\Gamma ~(1 + z)^2 \quad{\rm nT},
\label{eq:Blimit}
\end{equation}
where $\Gamma$ is the bulk Lorentz factor of the jet.  Even for
$\Gamma \approx 1$, at $z = 3.689$ any jet magnetic field less than 7
nT would cause iC losses on the CMB to dominate synchrotron
losses. The factor by which the iC losses are faster
  than those to synchrotron for a given electron is given by the ratio of
  energy density in the CMB, $u_{\rm CMB}$, to that in the magnetic
  field, $u_{\rm B}$.  For $B$ in units of nT,
\begin{equation}
u_{\rm CMB}/u_{\rm B} \approx  0.1 ~\Gamma^2 ~(1 + z)^4/B^2.
\label{eq:liferatio}
\end{equation}

Intrinsic minimum-energy magnetic fields of quasar kpc-scale jets are
difficult to measure because of uncertain relativistic beaming.  For
low-redshift quasar samples, values of a few to 10 nT are inferred
\citep[e.g.,][]{hogan11, marshall18}.  However, to model the beaming
such results assume that the X-ray emission in quasars at low redshift
is of iC origin, something now proved to be unlikely in general by
optical polarization measurements and gamma-ray upper limits for
several objects \citep[e.g.,][]{cara13, meyerg, meyer15, breiding}.
In bright kpc-scale jet knots of nearby powerful radio galaxies where
X-rays are detected and beaming is small and constrained by structural
measurements more reliably than for quasars, estimates from
synchrotron modelling typically find values of $B_{\rm me}$ between 10
and 30 nT \citep[e.g.,][]{wb05, kataoka, wfos}.  Unlike most quasars
included in Figure~\ref{fig:histo}, \source\ is at a redshift where
the relativistic electrons are expected to lose more of their energy
via iC scattering on the CMB (iC-CMB) than via synchrotron radiation,
even for modest values of $\Gamma$,

The measured
jet X-ray spectrum  of \source, of $\alpha_{\rm x} \approx 0.65$, is an
excellent match to that expected from synchrotron or iC radiation from
an electron spectrum that has undergone particle acceleration and not
suffered significant energy losses \citep[e.g.,][]{achterberg}.
Energy losses would be manifest as a steepening of the spectrum to
higher energies by a value of $\Delta\alpha \approx 0.5$ (or somewhat
larger). Such a break is commonly observed at rest frequencies between
$10^{12}$ and $10^{15}$ Hz in the kpc-scale jet knots of nearby radio
galaxies with detections spanning the radio and X-ray energy bands
\citep*[e.g.,][]{bm87, hard66b}, including jets of higher power and so
more comparable to quasars \citep[e.g.,][]{wb05, wfos}.  The
energy-loss time implied by the location of the spectral break often
roughly matches the light travel time across the knots, and so
particle acceleration is not required to occur throughout
the knots \citep[consistent with arguments made
    for Pictor-A based on short-timescale X-ray
    variability:][]{marshall-pica, hard-pica}, and the X-rays are
  undoubtedly of synchrotron origin in such sources.  Similar spectral
  breaks are seen in a number of low-redshift quasars, although the
spectra do not extrapolate to fit the X-ray data
\citep[e.g.,][]{sambruna06, godfrey12, breiding, meyer17}.  In the
high-redshift quasar \source\ it appears that acceleration processes
are not raising electrons to high enough energy for the spectrum to
reveal a signature of energy losses or for X-ray synchrotron emission
to be achieved,  and the high redshift provides
good reason to adopt an iC explanation for the X-ray emission. It is
further notable that the X-rays fade after the sharp radio bend
towards the end of the jet, consistent with an iC-CMB interpretation
for which the X-ray to radio ratio is more strongly dependent on angle
to the line of sight than alternative synchrotron explanations.

\subsection{Modelling results}
\label{sec:jetmodellingresults}%4

\begin{figure}
\centering
\includegraphics[width=0.75\columnwidth]{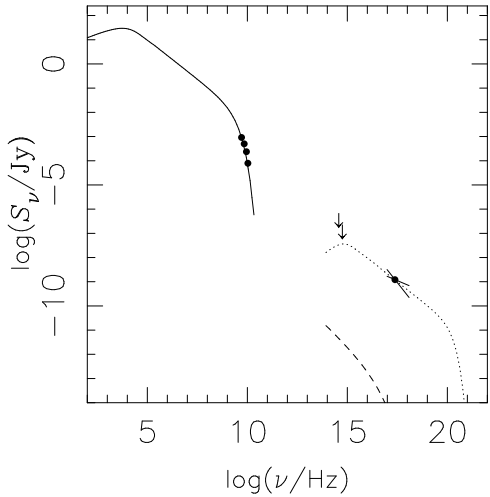}
\caption{Spectrum of region Jet D. Measurement detections and upper
  limits are from Tables~\ref{tab:xspec} and \ref{tab:fd}. Continuous line shows
  synchrotron emission from an electron spectrum modelled as described
in the text, with the radio data falling on the high-energy exponential
tail.  The dotted line is iC-CMB modelled to give the X-ray emission,
using parameters from Table~\ref{tab:modpars}, and the dashed line is
the synchrotron self-Compton prediction.  Plots for jet regions A, B,
and C are similar in their key features (Fig.~\ref{fig:appendix}).}
\label{fig:jetdsed}
\end{figure}

We have modelled the electron spectrum as extending from $\gamma_{\rm
  min} = 10$ to $\gamma_{\rm max}$ as
$\propto\gamma^{-2.3}$ so as to produce synchrotron emission of
spectral index $\alpha=0.65$ at radio frequencies below our
observations, and such that the X-rays match the spectrum expected
from iC scattering of the CMB. We used equations in the form given by
\citet{wrev} to model the X-rays via the iC-CMB process. An
illustrative example of the data and spectral model for the Jet-D
region is shown in Figure \ref{fig:jetdsed}.

%------------ begin figure
%------------ end figure

%------------ begin figure
\begin{figure*}
\centering
\includegraphics[width=0.5\columnwidth]{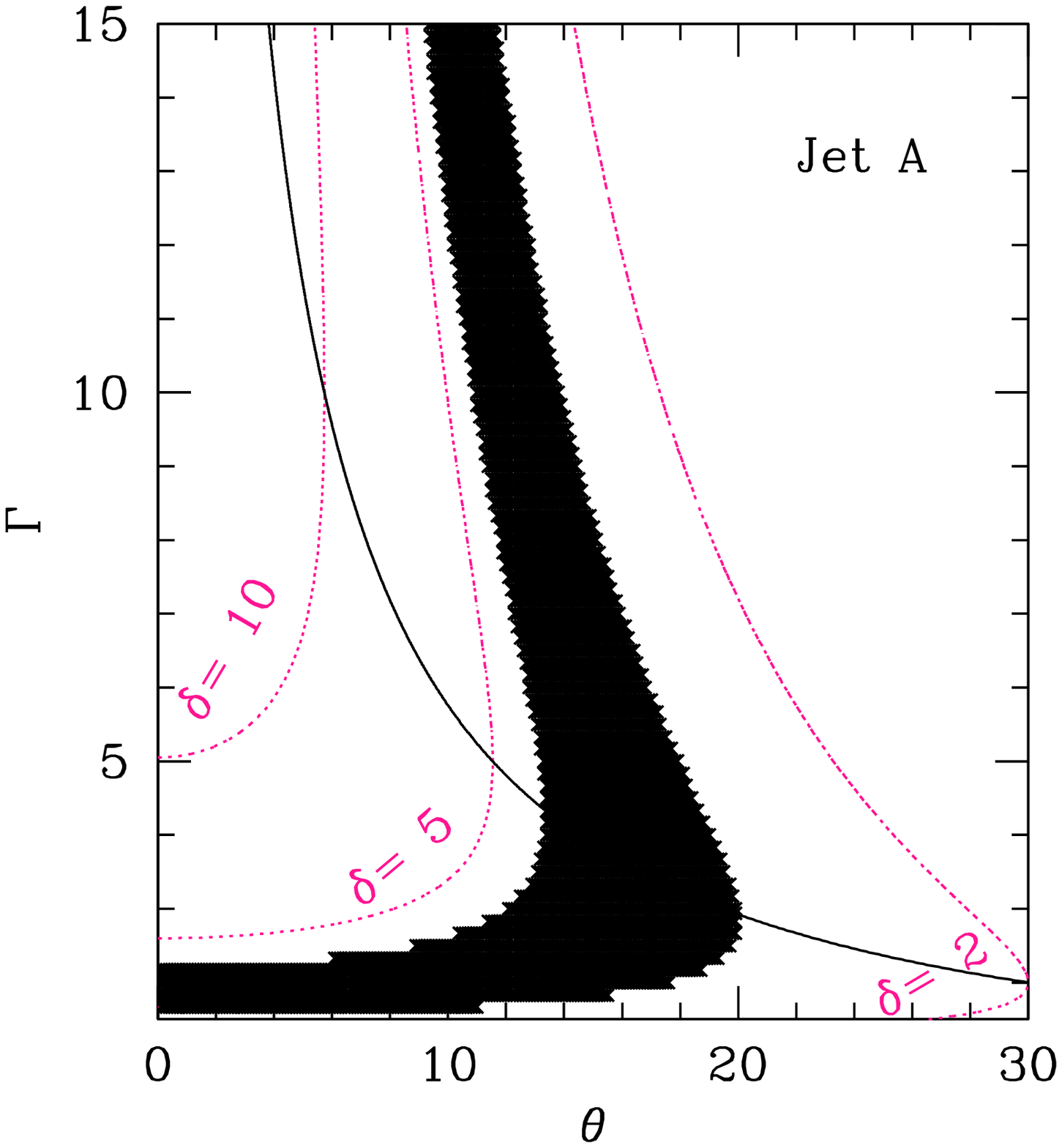}
\includegraphics[width=0.5\columnwidth]{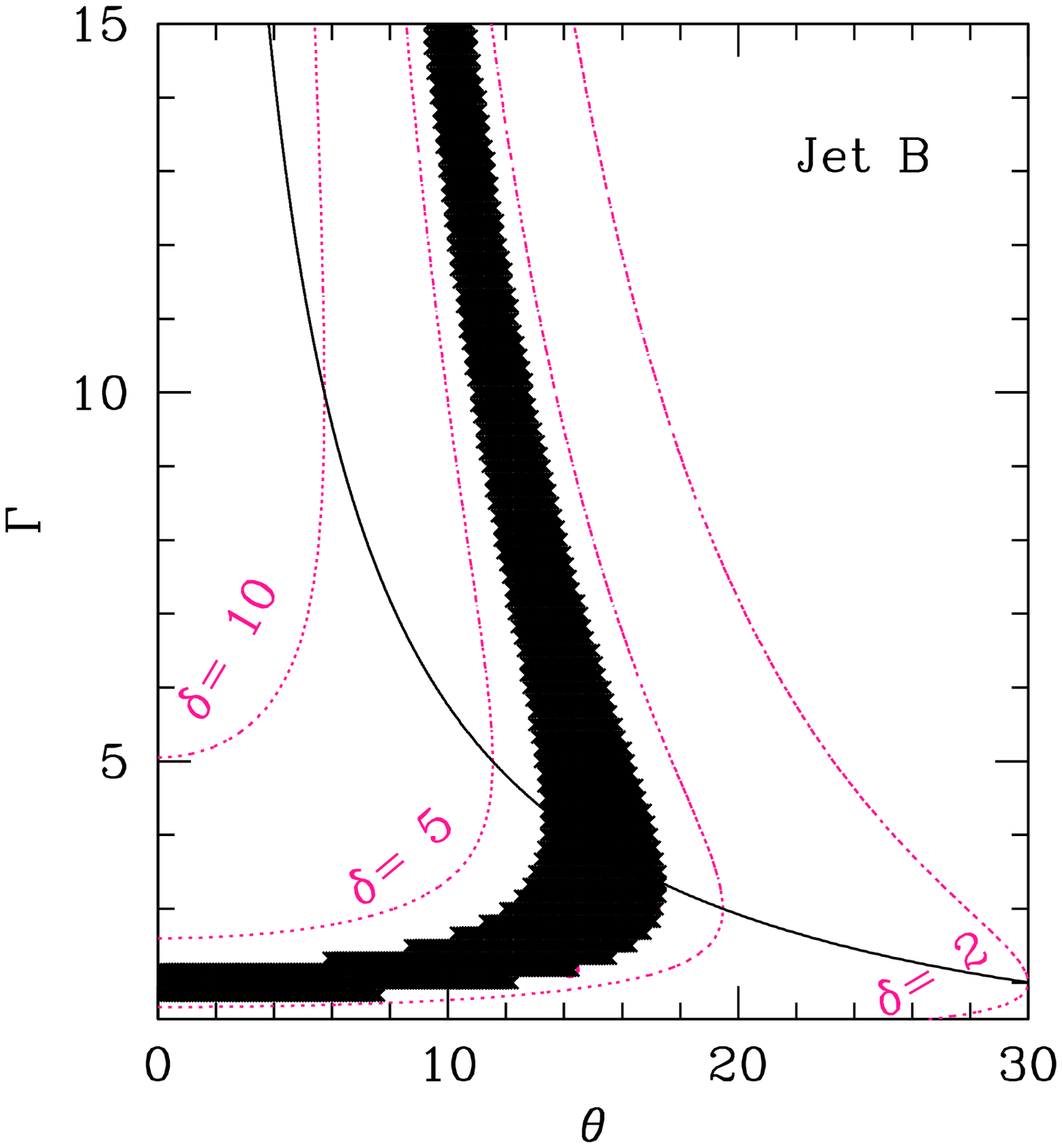}
\includegraphics[width=0.5\columnwidth]{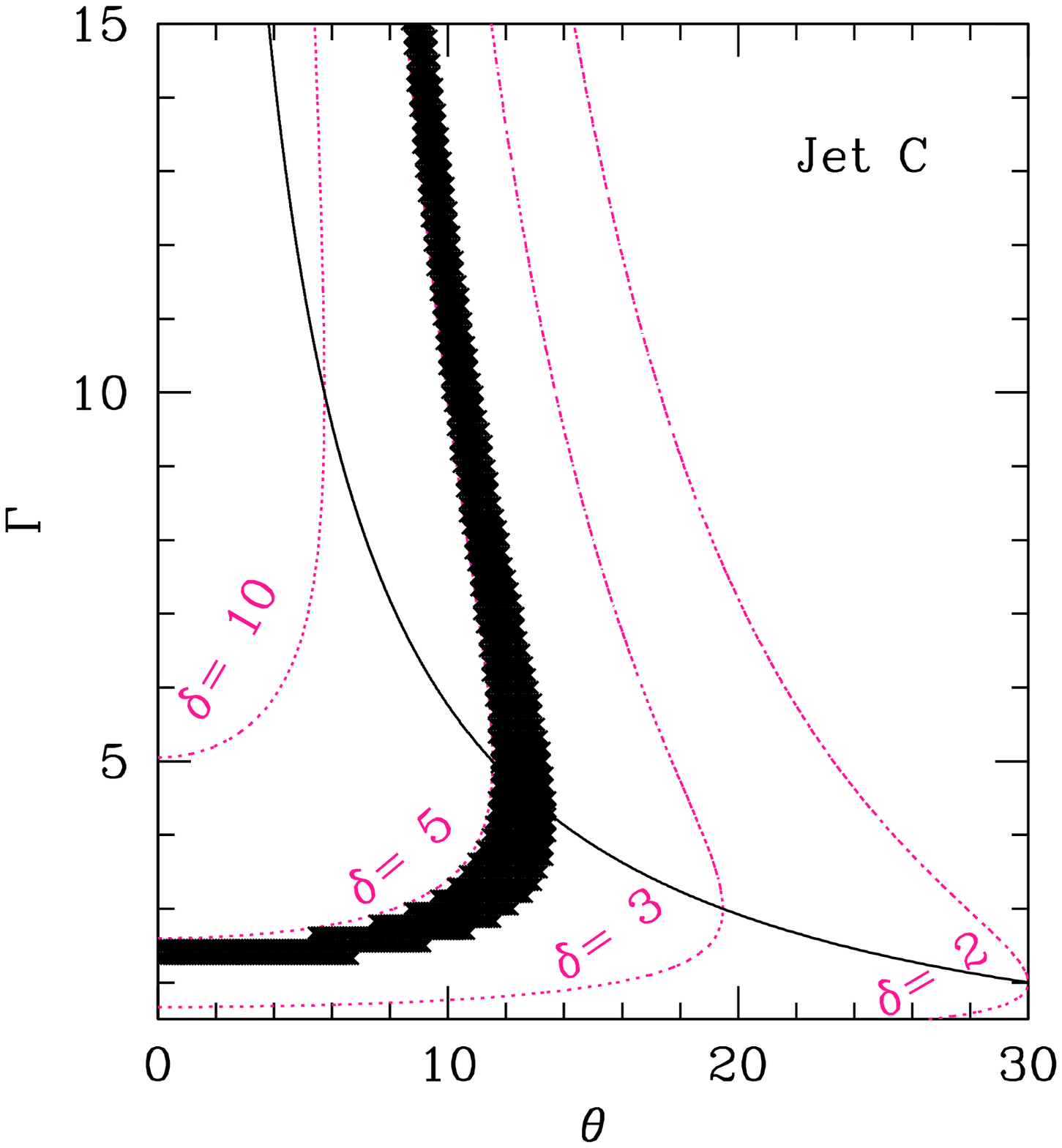}
\includegraphics[width=0.5\columnwidth]{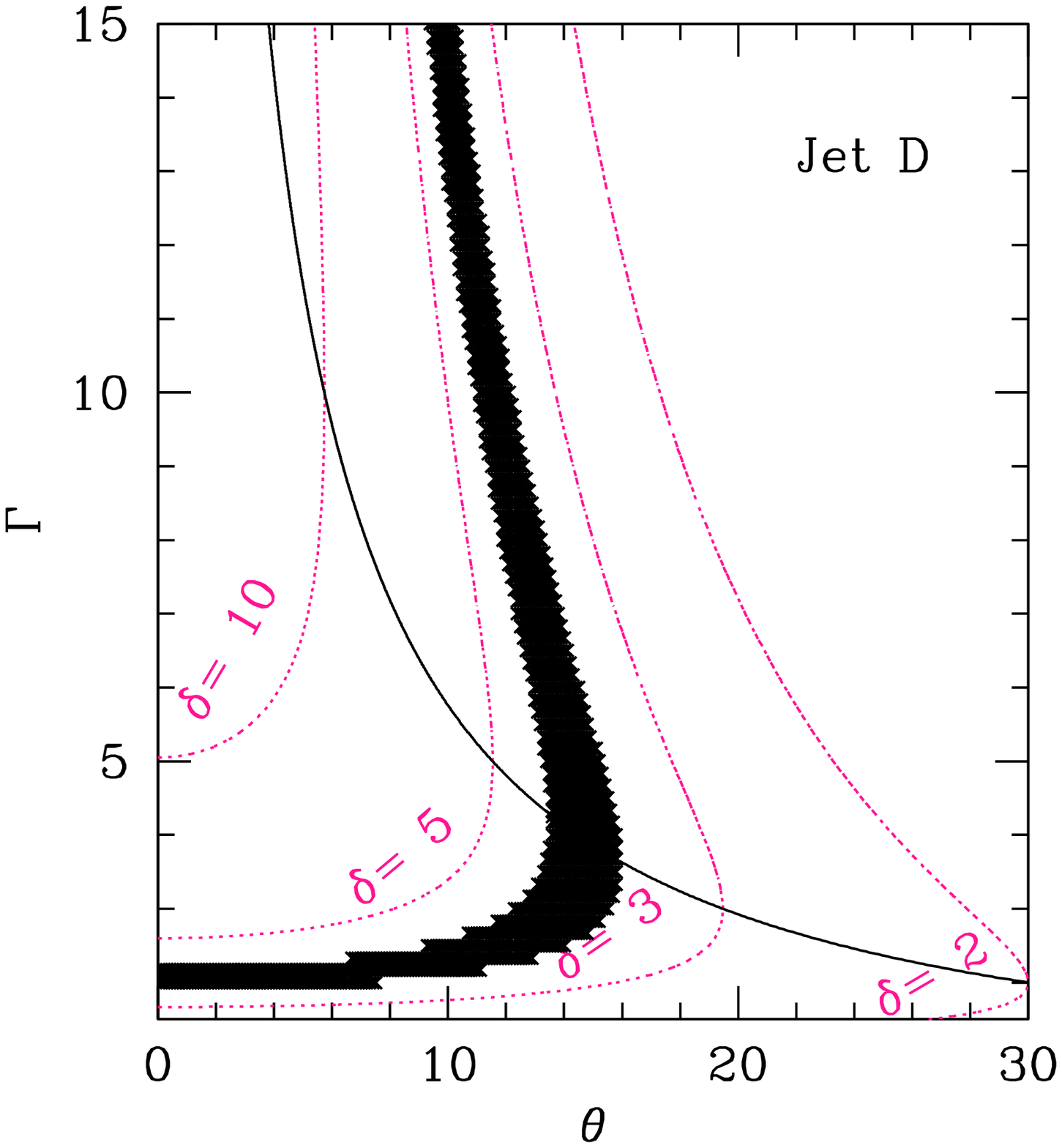}
\caption{ Lorentz factor and angle to line of sight for model fits to
  jet regions A to D. The shaded areas show allowed values of $\Gamma$ and
  $\theta$ within uncertainties on the X-ray measurements.  Dotted
  lines show contours of Doppler factor, $\delta$, and solid curves
  are the locus of $\delta = \Gamma$.}
\label{fig:jetbeampars}
\end{figure*}
%------------ end figure

%----------- begin table
\begin{table*}
\caption{Values for physical parameters, constrained to $\Gamma = 4$
  except in the case of $B$.}
\label{tab:modpars}
\renewcommand{\arraystretch}{1.3}
\begin{tabular}{lcccccccccc}
\hline
(1) & (2) & (3) & (4) & (5) & (6) & (7) & (8) & (9) & (10) & (11)\\
Region
 & $r$(arcsec)
  & $\gamma_{\rm max}$
  & $\Gamma$
  & $\theta$ (deg)
  & $\delta$
  & $\log_{\rm 10}(\tau_{\rm r})$ (yr)
  & $\log_{\rm 10}(\tau_{\rm x})$ (yr) 
  & $\nu_{\rm r}$ (MHz)
  & $B$ (nT) 
  & $P_{\rm jet} (10^{46}$ erg s$^{-1})$ \\
\hline
 Jet A  & 0.61 & 7200  & 4  & 15.3 & 3.8  & 4.6 & 6.0 & 1.6 & $3.09^{+1.1}_{-0.39}$&3.0\\
 Jet B  & 0.65 & 5000  & 4  & 14.8 & 3.9  & 4.8 & 6.0 & 1.5 &$2.66^{+0.57}_{-0.27}$&2.5\\
 Jet C  & 0.71 & 5000  & 4  & 12.3 & 4.6  & 4.8 & 6.0 & 1.0 &$2.31^{+0.26}_{-0.16}$&2.3\\
 Jet D  & 0.78 & 5000  & 4  & 14.5 & 4.0  & 4.8 & 6.0 & 1.7 &$2.95^{+0.39}_{-0.2}$ &4.5\\
\hline
\end{tabular}
\medskip
\begin{minipage}{\linewidth}
{ (1) Region from Fig.~\ref{fig:jetregions}. 
(2)  Angular radius of the equivalent spherical emitting region.  
(3) Maximum value of the electron Lorentz factor,
  as constrained by the steep radio spectra. 
(4) Value of jet Lorentz factor. 
(5) Angle to line of sight. 
(6) Doppler factor. 
(7) Logarithm of lifetime of radio-emitting electrons of  $\gamma_{\rm
    max}$. 
(8) Logarithm of lifetime of electrons emitting iC-CMB in the
  X-ray at 1 keV.  
(9) Frequency at which 1-keV-emitting electrons produce radio-synchrotron emission, in units of MHz.
(10) Intrinsic magnetic field strength with
  uncertainties that take into account the uncertainties in beaming
  parameters shown in Fig.~\ref{fig:jetbeampars}, and so are not constrained to $\Gamma = 4$. 
 (11) Jet power for $\Gamma = 4$ and cyclindrical section of radius given in column 2.
}
\end{minipage}
\end{table*}
%----------- end table

Angle to the line of sight, $\theta$, and bulk Lorentz factor,
$\Gamma$, give the Doppler factor, $\delta$.  Rather than assume
$\delta = \Gamma$, as has often been the case in quasar jet modelling
\citep[e.g.,][]{schwartz-fourjets}, here we explore all reasonable
values of $\Gamma$ and $\theta$ that give the observed level of X-ray
emission, treating each region as a separate
spherically-emitting volume.  Results are shown in
Figure~\ref{fig:jetbeampars}, where shaded regions indicate fits
within the X-ray normalization uncertainties of Table~\ref{tab:xspec}.
The {\it a priori\/} probability of the approaching jet
having a line-of-sight
angle between $\theta$ and $\theta + \Delta\theta$ is
$\sin\theta\Delta\theta$, and so large values of $\theta$ are
preferred, consistent with $\delta = \Gamma$.  This suggests that
$\Gamma$ is likely to have a value of a few, at most.  The
angle to line of sight is consistent with the range of values expected
for quasars, without needing to be of the very small value often
assumed true of a blazar.

In Table~\ref{tab:modpars} we give parameter values where $\Gamma = 4$
is adopted for all jet regions, under the hypothesis that the jet does
not suffer significant bulk deceleration.  The uncertainties given for
magnetic field strength in Table~\ref{tab:modpars} are not constrained
in this way, and they apply to all allowed combinations of $\Gamma$
and $\theta$ from Figure~\ref{fig:jetbeampars}. It is immediately
clear that for \source\ the condition in Equation (\ref{eq:Blimit}) is
satisfied, and from Equation (\ref{eq:liferatio})
the electron losses to iC scattering on the CMB are much faster
 (by a factor $>80$) than those to synchrotron
losses, including, of course, for those electrons
  responsible for the synchrotron radiation observed in the radio
  band.

With regards the dominance of iC losses we should mention an
additional uncertainty, which is jet volume.  
In Section
\ref{sec:imaging} we found that in general the jet is resolved in
width.  From Figure~\ref{fig:hm-trans} we find that the median transverse width
over the jet length incorporating regions A to D has a Gaussian $\sigma$  of 0.27 arcsec, for which the radius of
 an equivalent cylinder is 0.54 arcsec.  This is on average about 20 per cent smaller than the
radii used in  Table~\ref{tab:modpars}.
However, scaling with volume is relatively
weak.  For example, if the volume of jet region A is reduced by a
factor of four the black shaded region in Figure~\ref{fig:jetbeampars}
(left panel) moves a little to the left and up, to slightly larger
values of $\delta$.  The value of $B$ then increases slightly from
$3.1^{+1.1}_{-0.4}$ to $4.8^{+1.5}_{-0.6}$ nT, but not enough to affect
the conclusion of iC dominance.

The combination of the steeply falling radio spectrum
  and iC-CMB interpretation which requires only a single electron
  population, means that the jet should be largely invisible in the
  sub-mm to mid-infrared band and in high-energy gamma-rays
  (Fig.~\ref{fig:jetdsed}).

\subsection{Particle lifetimes}
\label{sec:jetmodellinglifetimes}%4

In Table~\ref{tab:modpars} we give values for the lifetimes of
electrons, using $\Gamma = 4$ as the example.  For the jet plasma to
reach region D it must travel a deprojected distance of about 130 kpc,
taking roughly $4 \times 10^5$~yr which is about six times longer than
the radiative lifetime of the electrons of $\gamma_{\rm max}$
producing the radio.  This argues for a number of regions of particle
re-acceleration along the jet, presumably related to the observed
non-uniform radio intensity and deviations from a straight path.
For iC losses dominating over synchrotron, as here,
  the lifetime of electrons of Lorentz factor $\gamma$ can be written
  simply as
\begin{equation}
\tau  \approx 2.3 \times 10^{12} \Gamma^{-2} \gamma^{-1} (1+z)^{-4} \quad{\rm yr}.
\label{eq:lifetime}
\end{equation}
An increase in $\tau$ by a factor of six would require $\Gamma \approx
1.6$ which Figure~\ref{fig:jetbeampars} shows to be inconsistent with
data for region C, and other regions would then prefer smaller
$\theta$ and so a longer jet. Values of $\Gamma > 4$ would require a
larger number of re-acceleration regions.  It is unlikely that the
requirement for re-acceleration along the jet can be removed.  It is
satisfying that the granularity in radio structure suggested by our
$\Gamma = 4$ solution is roughly consistent with the radio
observations, although it should be noted that the radio appearance is
strongly driven by radio beam size.

There is no requirement for re-acceleration along the jet for the
lower-energy X-ray-emitting electrons, and were jet fuelling to cease
we might expect eventually to detect this jet in the X-ray but not the
radio, before its ultimate disappearance.  We would first expect the
radio-loudness of the core to have diminished considerably.

An expectation from the lifetime of the
  X-ray-emitting electrons being longer than the jet travel distance
  is that the X-ray surface brightness should be smooth.
  \citet*{tav-clumps} highlighted the relatively strong correspondence
  of X-ray and radio knots in some low-redshift quasar jets as a
  problem for the iC-CMB model and sought solutions in terms of
  clumping and adiabatic losses.  In contrast, in \source\ the X-ray
  profile is flat beyond the core and out to 3 arcsec where there is
  brightening and sharp bending (Fig.~\ref{fig:hm-trans}), consistent
  with a long-lived electron population being responsible for the
  X-ray emission.  However, as the deconvolution
  (Fig.~\ref{fig:arestore}) suggests a degree of knottiness over this
  length, deeper data would be required to draw reliable conclusions.
  The range of surface brightness when the terminal region is included
  is a factor of roughly three which is less than the factor greater
  than 10 typical of low-redshift quasar X-ray jets such as in 3C\,273
  \citep{sambruna3c273, marshall3c273} and PKS\,0637-752
  \citep{schwartz0637, chartas0637} and closer to expectations for the
  iC-CMB model.  We note that \source\ is significantly shorter in
  angular extent and presents fewer resolution elements than the
  low-redshift jets for which more-detailed multiwavelength
  comparisons are possible.

\subsection{Jet power}
\label{sec:jetmodellingpower}%4

For a tangled minimum-energy magnetic field of energy density
$u_{\rm B_{\rm me}}$, where minimum energy is calculated
 with respect to all particles and the field, we can re-cast equation B17 of
\citet{schwartz-fourjets} to give jet power, for $\alpha \ne 0.5$, as
\begin{equation}
\label{eq:pjet}
\begin{split}
 P_{\rm jet} \approx & {\Gamma^2 \beta c A u_{\rm B_{\rm me}} \over (1+\alpha)}\Bigg[   
  {4 (3+\alpha) \over 3}  + \\
  & {(\Gamma- 1) \over \gamma_{\rm min}\Gamma}{(1+k_2)\over (1+k_1)} 
   \left(2\alpha -1 \over 2\alpha\right) 
  \left[ {1 - (\gamma_{\rm max} /\gamma_{\rm \min})^{-2\alpha} \over 
   1 - (\gamma_{\rm max} /\gamma_{\rm min})^{-(2\alpha-1)}}\right]\Bigg].
\end{split}
\end{equation}
 Here $\beta c$ is jet velocity ($\beta=\sqrt{1-\Gamma^{-2}}$), $A$ is
 the cross-sectional area of the jet, $\alpha, \gamma_{\rm min},
 \gamma_{\rm max}$ are the spectral parameters of the
 synchrotron-emitting component as defined in
 Section~\ref{sec:jetmodellingradio}, $k_1$ is the ratio of the energy
 density of other particles to that of the synchrotron-radiation
 emitting electrons/positrons, and $k_2$ is the ratio of rest-mass
 energy of other particles to that in electrons/positrons.

Our calculations in Table~\ref{tab:modpars} are for an
electron--positron jet, with $k_1 = k_2 = 0$, for reasons given in
Section~\ref{sec:jetmodellingradio} \citep*[and see fuller discussion
  in][]{sikora}.  The average $P_{\rm jet}$ for
  the four regions is roughly $3 \times 10^{46}$ erg s$^{-1}$, and
  the spread of values (Table~\ref{tab:modpars}) is indicative of
  systematic uncertainties given the choice of beaming parameters.

The jet power we have found is small in comparison
  with the quasar luminosity of about $4 \times10^{47}$ erg s$^{-1}$ 
 (Section~\ref{sec:core}), and we consider
  if it might be underestimated.  A simple way of allowing a larger
  value of $P_{\rm jet}$ is to select larger $\Gamma$ consistent with
  Figure~\ref{tab:modpars}. Very large $\Gamma$ would make
  \source\ highly special, and so we adopt $\Gamma = 10$ as a likely
  upper bound, giving a new average upper bound over the four regions of
  $P_{\rm jet} \approx 2 \times 10^{47}$ erg s$^{-1}$.

The other potential contributor to higher jet
  power would be heavy particles in the jet.  In
  Equation~\ref{eq:pjet}, the first term represents the rate of
  advection of internal energy (particles and magnetic field), and is
  largely independent of particle composition, except through a small
  positive increase of $u_{\rm B_{me}}$ with increasing $k_1$.  The
  second term, which represents the contribution due to the ram
  pressure of the particles, is negligible compared with the first
  term in the case of an electron--positron jet, but becomes dominant
  with the insertion of heavy particles due to the increase in $k_2$.

  An extreme case is to allow charge neutrality by pairing electrons
  with positive ions. If the positively-charged material has cosmic
  abundances, then $k_2 \approx 2200$ \citep{schwartz-fourjets}.  The
  result for $P_{\rm jet}$ then depends on how energetically important
  the positive ions are compared with the electrons, and the smaller
  the value of $k_1$ the larger the contribution to $P_{\rm jet}$ (the
  weak positive dependence of $u_{\rm B_{me}}$ on $k_1$ is more than
  offset by the $(1+k_1)^{-1}$ factor).  Were charge neutrality
  preserved using a proton spectrum of shape similar to that of the
  electrons, the minimum Lorentz factor of the protons would be
  $k_1/k_2$ that of the electrons, which is a situation that is
  violated for low $k_1$ and the low value of electron $\gamma_{\rm
    min}$ needed for a iC-CMB model to explain X-ray emission.  Thus,
  to allow both charge neutrality and a low $k_1$ \cite[a value $k_1 =
    1$ is used, for example, by ][]{schwartz-fourjets} the proton
  spectrum must either be much flatter than that of the electrons, and
  not easily matched to particle acceleration models, or the proton
  spectrum must turn up rapidly at low energies such that most of the
  jet protons are cold.  Although we consider charge neutrality
  provided by positive ions to be unlikely, in the situation of $k_1 =
  1, k_2 = 2200$, we find for \source\ that $P_{\rm jet} \approx 2
  \times 10^{47}$ erg s$^{-1}$ for $\Gamma = 4$, and it would reach
  about $10^{48}$ erg s$^{-1}$ for $\Gamma = 10$.  These last
  two jet powers increase by a factor of roughly two if minimum-energy
  is between only the electrons and field despite the presence of
  heavy particles (a disappearance of the $(1+k_1)^{-1}$ term on the
  right hand side of Equation~\ref{eq:pjet} dominates that result).

Our discussion has shown that quasar jet power is uncertain.  In what
we consider to be the most likely situation of an electron--positron
jet, and adopting $\Gamma \approx 4$, the power is
  about $3 \times 10^{46}$ erg s$^{-1}$.  It could potentially reach
  $10^{48}$ erg s$^{-1}$ in the extreme situation of the jet being
  very fast, gaining charge neutrality from positive ions, and
  minimum energy applying between the light particles and magnetic
  field alone, but jet power would then exceed estimates of the quasar
  bolometric luminosity.

In Section~\ref{sec:jetspec} we found the 2--20~keV X-ray luminosity
of the jet to be $3.2 \times 10^{45}$ erg s$^{-1}$, under isotropic
assumptions, giving a modelled bolometric luminosity
  between observed frequencies of $10^{15}$ and $10^{20}$ Hz
  (see Figures~\ref{fig:jetdsed} and \ref{fig:appendix}) a factor of nine
  larger.  For iC-CMB emission we must take into account the beaming
pattern, which causes a correction of $\delta^{-(4+2\alpha)}$
\citep{dermer}. This corrects the isotropic X-ray
  luminosity to be about $2 \times 10^{42}$ erg s$^{-1}$ and the
  bolometric jet luminosity to be $L_{\rm bol-jet} \approx 2 \times
  10^{43}$ erg s$^{-1}$. This is only 0.07 per cent of the jet power
  of $P_{\rm jet} \approx 3 \times 10^{46}$ erg s$^{-1}$ for an
  electron--positron jet.  The radiated fraction is smaller if the jet
  contains heavy particles. Essentially all the jet power is therefore
  available for heating ambient gas.

\section{The quasar core}
\label{sec:core}%5

\begin{figure}
\centering
\includegraphics[width=0.55\columnwidth]{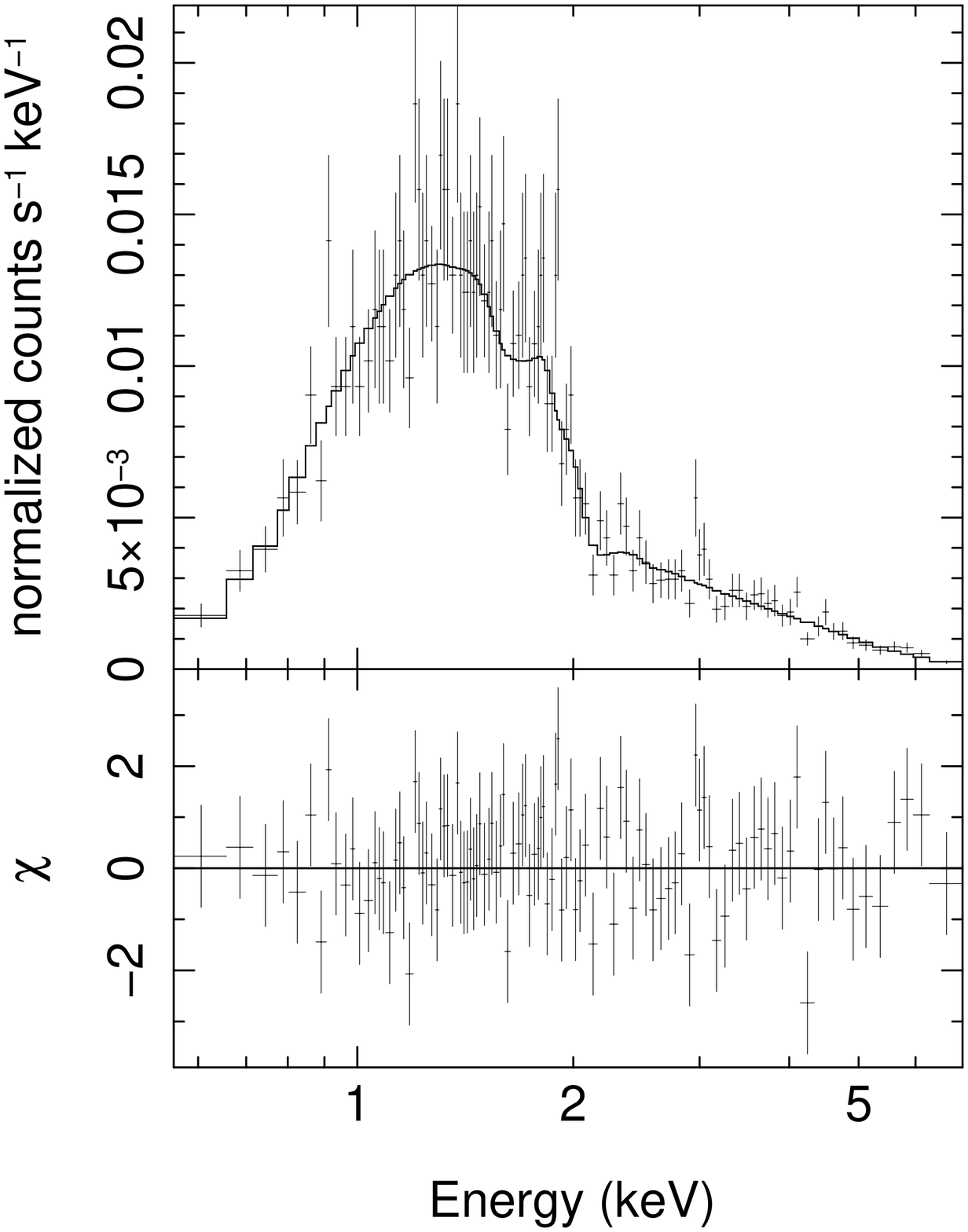}
\includegraphics[width=0.40\columnwidth]{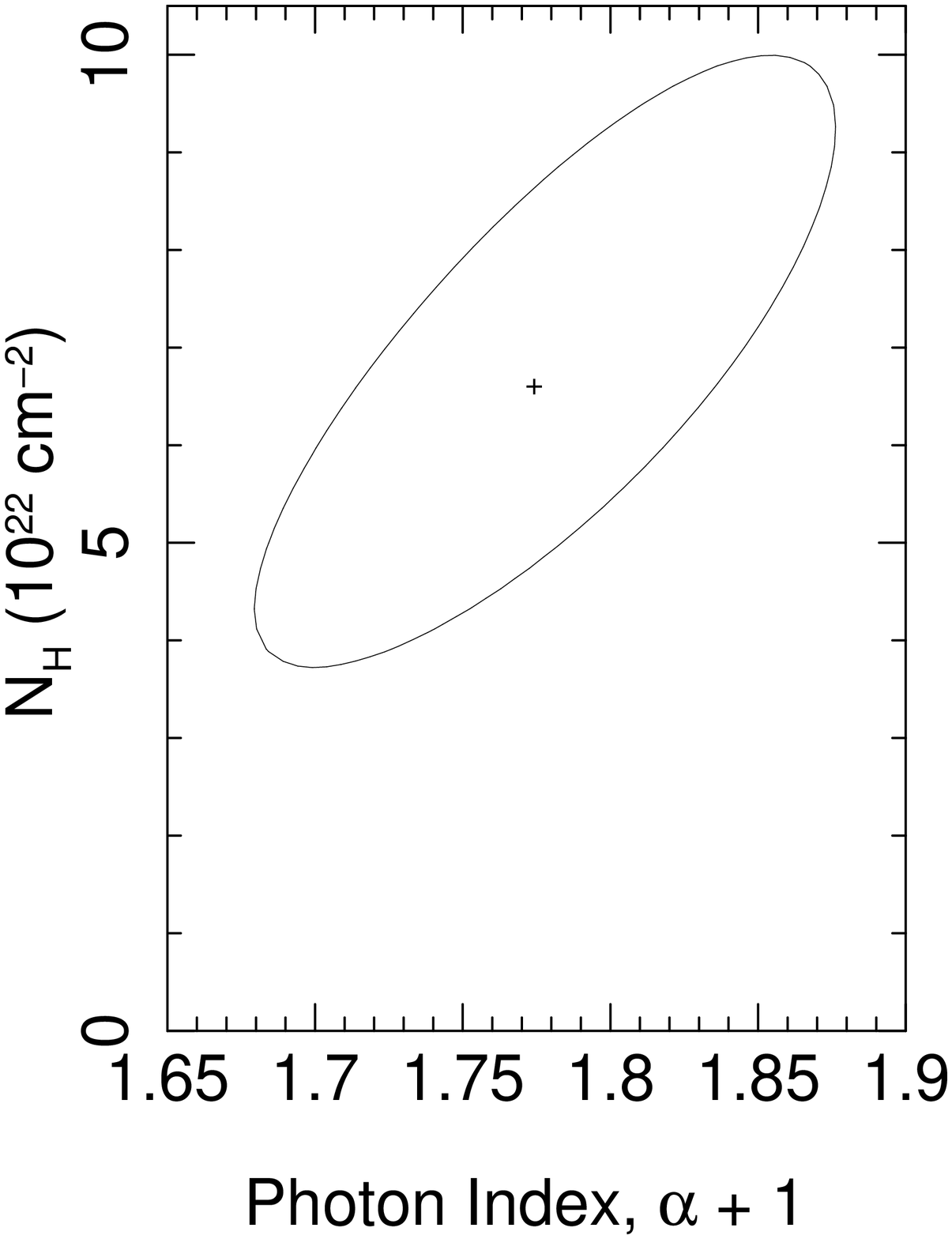}
\caption{ Left: X-ray spectrum of the core fitted to a power law and
requiring additional absorption at the redshift of \source\ (Table~\ref{tab:xspec}).
Right: Uncertainty contour in spectral index and intrinsic $N_{\rm H}$ shown at the
level of 90\% confidence for one interesting parameter.}
\label{fig:corespec}
\end{figure}

The X-ray spectrum of the core gives a good fit to a single-component
power law but there is relatively high absorption in excess of the
Galactic value (Figure~\ref{fig:corespec}).  Visual comparison between
the shape of the jet spectrum (Figure~\ref{fig:jetspec}) and that of
the core (Figure~\ref{fig:corespec}) shows it is indeed in their
relative levels of low-energy emission where they differ.

%------------ begin figure
%------------ end figure

When the excess absorption is placed as intrinsic to the source, the
column density is $6.7^{+3.4}_{-2.9} \times 10^{22}$ cm$^{-2}$.  The
spectral index, $\alpha_{\rm x} = 0.78\pm0.10$, for this high-redshift
source does not follow the trend hinted at in \citet{marshall18} of
core spectral indices becoming flatter with increasing redshift.
Details of the spectral fit are given in Table~\ref{tab:xspec}.  The
source-frame luminosity, corrected for absorption, is $3.0 \times
10^{46}$ erg s$^{-1}$ (2--20 keV).

The redshift of the excess absorption is not constrained by the data,
with any redshift between 0 and 4 lying within 90\% confidence bounds.
We could find no reported evidence of absorption in our Galaxy in this
relatively high-Galactic-latitude direction that would not be taken
into account by the Galactic value, for example molecular gas,
 and it would be highly unusual for that gas to
  have a lot of structure on a scale of a few arcsec, so as to affect
  the core and not the jet. (Formally we can set a 90 per cent upper
  limit of excess absorption $N_{\rm H} < 2.6 \times 10^{21}$
  cm$^{-2}$ at $z=0$ in front of the jet.)  Although it is notable
that the quasar is associated with a damped Ly$\alpha$ (DLA) system at
$z=3.449$, the DLA profile suggests N$_{\rm H\,I} = 2.5 \times
10^{20}$ cm$^{-2}$ \citep{ellison01}.  While
  \citet{bechtold} have detected X-ray absorption of order $10^{21}$
  cm$^{-2}$ potentially due to a DLA system for one quasar, for
\source\ absorption attributable to the DLA is significantly lower
than the X-ray absorption we find.  So, the X-ray absorption of 
$\approx 7 \times 10^{22}$ cm$^{-2}$ is likely to be 
intrinsic to the quasar.  \citet{ellison05} note a DLA- and
X-ray-derived discrepancy for another quasar, B038-436 at $z=2.863$
(DLA at $z=2.347$), pointing out that dustless gas near the quasar or
an extinction curve significantly different from that of our Galaxy or
the Small Magellanic Cloud might be responsible.  The discrepancy is
of similar size for \source, and the same explanations may apply.
Such high intrinsic absorption is unexpected from simple unification
schemes \citep[e.g.,][]{barthel}.

Indeed, it has been known for decades that the cores of radio-loud
quasars, can, but do not always, show evidence of excess intrinsic
absorption, and it has been claimed that the level of that absorption
increases with increasing redshift \citep[e.g.,][]{elvis,reeves,
  page}.  There are known systematic effects in the sense that
absorption shows up first at the lowest X-ray energies, and so for a
fixed observing band the absorption must be larger for detection in
higher-redshift sources, complicating statistical inference related to
redshift trends.  While it has been suggested that
  intrinsic spectral curvature might instead be responsible for appearing as
  excess absorption \citep[e.g.,][]{tavecchioRBS315}, and deciphering the
  correct model is difficult \citep*{benhaim},
  most authors favour absorption along the line of sight, possibly
  associated with a warm-hot intergalactic medium \citep[e.g.,][]{eitan, arcodia}.
  Where absorption is detected in
high-redshift sources it has tended to be at levels of a few times
$10^{22}$ cm$^{-2}$ \citep{yuan2006}, and this is true for \source.
We summarize properties for the high-redshift quasars with resolved
X-ray jets from Table~\ref{tab:highz} in Table~\ref{tab:highzcorejet},
where it can be seen in column (3) that three of the five have such
levels of intrinsic absorption, while two do not.

%----------- begin table
\begin{table*}
\caption{Core and jet properties of quasars at $z > 3.5$ with X-ray jets, listed in order of increasing $z$.}
\label{tab:highzcorejet}
\setlength\tabcolsep{3.0pt}
\renewcommand{\arraystretch}{1.3}
\begin{tabular}{lccccccccl}
\hline
(1) & (2) & (3) & (4) & (5) & (6)& (7) &(8) & (9) & (10)\\
Name
  & $L_{\rm core_X}$ (2--10 keV) 
  & $N_{\rm H_{core(int)}}$
  & $\alpha_{\rm core_X}$
  & $L_{\rm jet_X}$ (2--10 keV) 
  & $L_{\rm jet_X}/L_{\rm core_X}$
  & $\alpha_{\rm jet_r}$
  & $\alpha_{\rm jet_X}$
  & Jet
  & Ref. \\
  & $10^{46}$ erg s$^{-1}$
  & $10^{22}$ cm$^{-2}$
  & 
  & $10^{45}$ erg s$^{-1}$
  & $\%$
  &
  &
  & $\nu_{\rm X}S_{\rm X}/\nu_{\rm
  r}S_{\rm r}$
  &  \\
\hline
PMN J2219-279 & 1.8 & $< 0.6$ & $0.34 \pm 0.06$ & 0.4 & 2 & ... & ...& ...& 1\\
\source\  & 2.0 & $6.7^{+3.4}_{-2.9}$ & $0.78\pm 0.1$ & 2.1 &10& Table~\ref{tab:fd} &$0.65\pm0.19$& 71 & 2\\
4C+62.29  & 5.7 &  $2.35^{+0.78}_{-0.71}$ & $0.74^{+0.08}_{-0.14}$ & 4.2&7& $0.93 \pm 0.09$ &$0.62^{+0.16}_{-0.17}$&22 & 3\\
GB 1508+5714 & 2.8 & $< 0.33$ & $0.52 \pm 0.04$ & 0.7 &2.5 & $1.4 \pm 0.2$ &$0.83^{+0.15}_{-0.14}$& 158 & 4, 5, 6, 7\\
GB 1428+4217 & 40 & $2.32^{+1.29}_{-1.26}$ & $0.44\pm 0.04$&1.3& 0.3 & $1.0 \pm 0.1$ &$0.64^{+0.39}_{-0.38}$& 205 & 6, 7\\
\hline
\end{tabular}
\begin{minipage}{\linewidth}
(2) Core unabsorbed rest-frame X-ray luminosity; 
(3) Core X-ray intrinsic column density; 
(4) Core X-ray spectral index; 
(5) Jet isotropic rest-frame X-ray luminosity; 
(6) Jet/core X-ray ratio, where both treated as isotropic;
(7) Jet radio spectral index, where available; 
(8) Jet X-ray spectral index, where available; 
(9) Jet X-ray to radio ratio, where available, using flux densities at 1 keV and 5 or 1.4~GHz;
(10) References: 
1. \citet{saez}; 
2. This work;
3. \citet{cheung1745}
4. \citet{siemiginowska1508}
5. \citet{yuan2006}
6. \citet{mckeough}
7. \citet{cheung1428}
\end{minipage}
\medskip
\end{table*}
%----------- end table

Our \hst\ data give quasar flux densities of $134\pm 3$ and 
$82\pm 3$ $\mu$Jy at $3.725 \times 10^{14}$ and $5.65 \times 10^{14}$ Hz,
respectively, confirming the level of redness reported by
\citet{ellison05}. Even the lower-frequency band is
  sampling emission from a rest-frame wavelength less than 250 nm for
  which the spectral-energy-distribution fitting of \citet{elvisatlas} for
  lower-redshift quasars gives an average bolometric correction of roughly a
  factor of six.  If we are to apply this to \source\ we find a
  bolometric quasar luminosity of $L_{\rm bol} \approx 4 \times 10^{47}$ erg
  s$^{-1}$, and using the expression in \citet{schwartz-fourjets} the energy
  density of quasar emission exceeds that of the CMB only within about 13
  kpc from the core, or one tenth the distance down the jet
  based on modelling in Table~\ref{tab:modpars}.  We are thus justified in
  ignoring this additional radiation field in Section~\ref{sec:jetmodellingresults}.
 
\begin{figure}
\centering
\includegraphics[width=0.85\columnwidth]{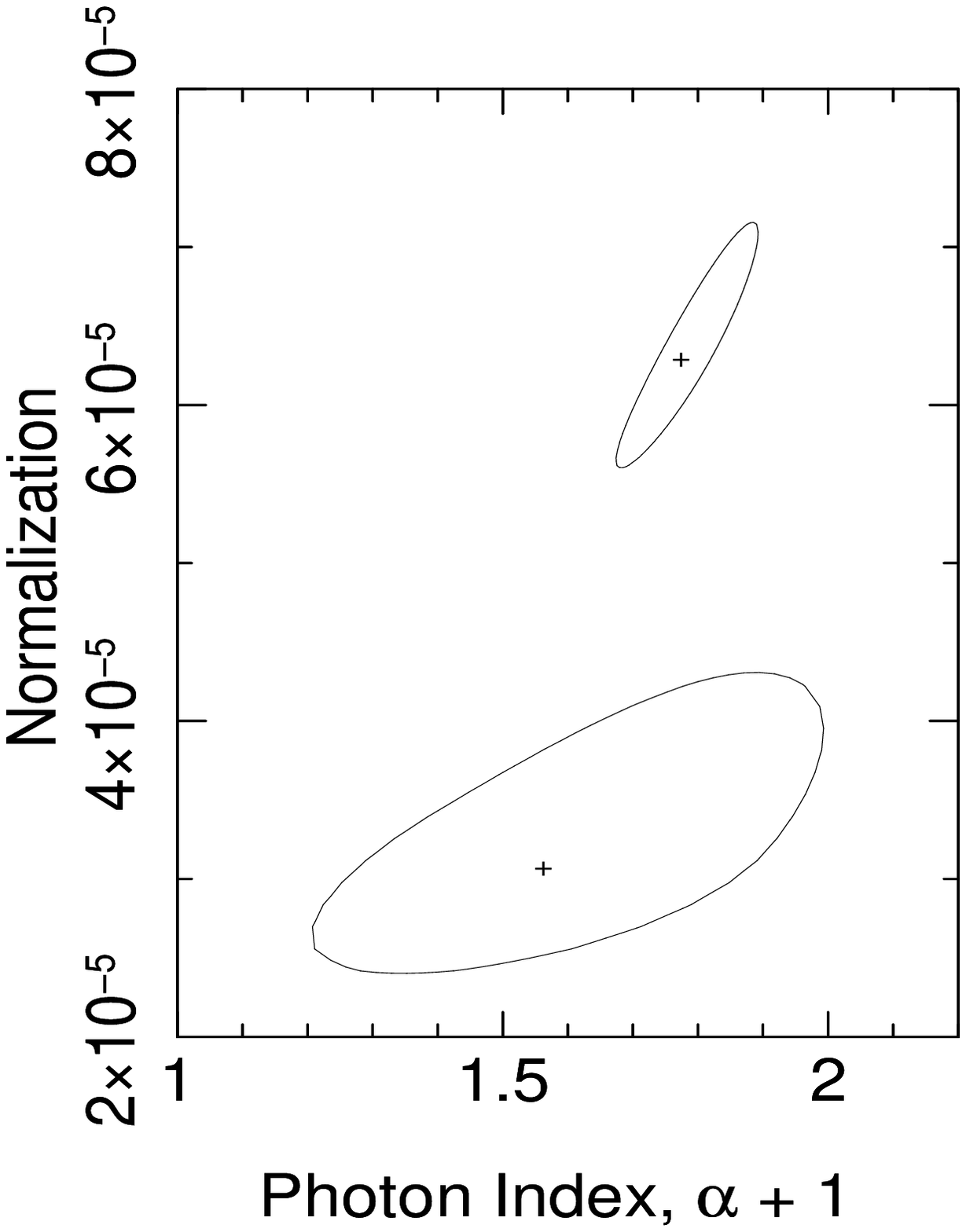}
\caption{Uncertainty contours (90\% confidence for one interesting parameter)
in spectral index and 1~keV-normalization
(photons cm$^{-2}$ s$^{-1}$ keV$^{-1}$) for the core emission from
data taken in 2018 (upper contour) and the archival data from 2007 (lower
contour). The results show variability of about a factor of two over
eleven years.}
\label{fig:corevar}
\end{figure}

The X-ray results are heavily dominated by the longer 2018 data (see
Table~\ref{tab:chandraobs} for dates).  In order to search for
possible variability between 2007 and 2018 we must use model-fitted
results because of changes in the \chandra\ response, and we have
fitted the power-law model with intrinsic absorption separately to the
data from the two epochs.  The (poorly constrained) absorption
measured from 2007 is consistent within errors with that from 2018,
but the normalization indicates significant brightening (by a factor
of about two) over the eleven-year interval (Fig.~\ref{fig:corevar}).
Within the 2018 observing period it suffices to examine count rates,
and no day-to-day variations are measured over the five days of
observation.  The long-term variability translates into an upper limit
on source size of about $11/(1+z)$ light years, or roughly 0.7~pc.

%------------ begin figure
%------------ end figure

\section{High-$z$ quasars with resolved X-ray jets}
\label{sec:summaryhighz}%5

Table~\ref{tab:highzcorejet} compiles interesting core and jet
properties of \source\ alongside those of the four other quasars at $z
> 3.5$ with resolved X-ray jets listed in Table~\ref{tab:highz}.
We see from the core emission that they are all relatively
X-ray-luminous quasars.  In contrast, out of a sample of 15 quasars at
$z > 4.5$ selected by compact steep-spectrum radio emission, only two
have rest-frame 2--10~keV luminosity more than $10^{46}$ erg s$^{-1}$
\citep{snios}. An uncertain but significant fraction of the unresolved
X-ray emission in the X-ray-jetted quasars is likely to be beamed and
related to pc-scale jet emission, contributing to enhanced luminosity
and the relative flatness of the X-ray spectrum, as historically
argued for lower-redshift samples \citep{zamorani, worrall87,
  wilkeselvis87} and more recently studied at high redshift
\citep[e.g.,][]{, bassett, wu}.  Given that
  variability has set an upper limit to the size of the nuclear X-ray emission
  region of about 0.7 pc in \source, a radio study using Very Long Baseline
  Interferometry would be a crucial to modelling that
  could estimate if the fraction of core X-ray emission attributable
  to a (misaligned) blazar-type jet is significant in this
  quasar.

The jet luminosities in
  Table~\ref{tab:highzcorejet} are calculated assuming the X-ray
  emission is isotropic, but this is highly unlikely to be the case
  whatever the true emission mechanism.  In
  Section~\ref{sec:jetmodellingpower} we comment on the strong
  correction if indeed an iC-CMB explanation holds.  The (isotropic)
  jet-to-core ratio (column 6) does however give a value that can be
  easily compared between sources.  We note that the value of 10 per
  cent for \source\ is high, which may be related to some combination
  of favourable beaming and weaker central AGN.  There does not appear
  to be a trend in jet-to-core ratio with redshift.  For
the 30 quasars at $0.2 < z < 2.2$ with detected X-ray jets in the
sample of \citet{marshall18}, the mean and median for this ratio are
3.8 and 2 per cent, respectively. Three sources have a jet-to-core
ratio above $10$ per cent.  These are B1421-490
  ($z=0.662$), B1202-262 ($z=0.789$) and B1030-357 ($z=1.455$), all
  with relatively bright terminal hotspots in X-ray and radio.  While
  the large value of X-ray jet-to-core ratio in \source\ is unusual,
  the large spread of values over sources does not appear to lend useful diagnostic
  power related to emission mechanisms to this parameter.

Particularly notable about \source\ is its exponentially falling jet
radio spectrum.  While jet radio spectral information has been published
only for three of the four other quasars, we note from column (7) of
Table~\ref{tab:highzcorejet} that they are all steep compared with
typical quasar cores.  If those jets are similar to \source\ in having
a maximum Lorentz factor with a value in the thousands, their radio
emission should also plummet at higher frequencies
  --- something that could be tested with higher-frequency radio
  data.

Column (8) of Table~\ref{tab:highzcorejet} shows
  that all the sources have jet X-ray spectra consistent with emission
  from particles that are newly accelerated and have not suffered
  significant energy losses.  This is accommodated easily within an
  iC-CMB framework, because low-energy electrons are producing the
  emission.  It is unfortunate that such electrons produce their
  synchrotron emission at very low MHz frequencies, inaccessible to direct
  observation (see, for example, column (9) of
  Table~\ref{tab:modpars}.

Column (9) of Table~\ref{tab:highzcorejet} gives the jet ratio of
monochromatic observer-frame flux density times frequency in the X-ray
and radio, as indicative of the relative output in the two bands.
The range of only one order of magnitude in this parameter might be
taken as indicating a close relationship between
the emissions, as is true for an iC-CMB model which relates them through a single electron
population, albeit extrapolated to different spectral energies.   Equation (8) of
\citet{wrev} shows that for the iC-CMB model we expect
\begin{equation}
{\nu_{\rm X}S_{\rm X} \over \nu_{\rm
  r}S_{\rm r}}
\propto {\delta^{1+\alpha}\over
B^{1+\alpha}} (1 + z)^{3 + \alpha} \left({1 + \cos\theta \over 1 +
\beta}\right)^{1+\alpha}.
\label{eq:beamic}
\end{equation}
The range of values in Table~\ref{tab:highzcorejet} can easily be accommodated by a
  combination of the increasing values of $z$ (see Table~\ref{tab:highz}) and small
  variations in beaming factors and intrinsic magnetic field strength.
  With large samples it should be possible to use high-redshift sources to construct expectations
  of the contribution of the iC-CMB mechanism at low redshift and compare with
  results there, particularly for cases where the iC-CMB mechanism
  appears ruled out as the main emitter of the X-rays and so predictions should fall below observations.

\section{Summary}
\label{sec:summary}%6

We have presented new results from \chandra,
  \hst\ and the \vla\ for the $z=3.69$ radio-loud quasar \source,
  studying the core and extended jet.  We find very steep radio
  spectra in the extended jet allowing, for the first time in an
  X-ray-jetted quasar, a maximum Lorentz factor for the
  radio-synchrotron-emitting electrons to be established.
  High-resolution radio mapping at lower frequencies would test the
  predicted spectral flattening to frequencies where the spectrum is expected to
  reflect that from particle acceleration.

PKS J1421-0643's high redshift makes it a prime candidate for
application of the iC-CMB model to the \chandra-detected X-ray
emission, extending about 4.5 arcsec from the core.
The measurement of an X-ray spectral index typical
  of radiation from newly accelerated particles is consistent with
  such an interpretation.  We have demonstrated that the data follow
the expectations of the model for a modest magnetic field strength
of a few nT, Doppler factor of about 4, and viewing angle of about
15$^\circ$.  Such parameters imply that the jet extends for about 130
kpc.  Under the model, the electrons emitting the highest-energy radio
emission need a few separate acceleration regions along the jet,
roughly consistent with the level of knottiness and bending that is
seen.

Emission plummets in the sub-mm to mid-infrared bands,
  and at high-energy gamma rays above about $10^{20}$ Hz, under such an
  iC-CMB description if there is no additional ultra-relativistic 
  particle population.
  This creates difficulties for methods used to rule out iC-CMB as the
  dominant X-ray emission mechanism of some quasar jets at low
  redshift.  We have compared measurements of \source\ with those of the
  four other $z > 3.5$ quasars with resolved X-ray jets.  With
  sufficiently large samples it should be possible to use modelling at
  high redshift to predict the contribution of iC-CMB to the X-rays of
  quasars at low redshift.  Such predictions would be expected to be
  below detected levels in some quasars.

We have discussed the considerable uncertainties in
  estimating jet power from uncertain jet composition and
  minimum-energy conditions.  If we rely on converging evidence
  favouring an electron--positron jet, the preferred jet
  power is of order $3 \times 10^{46}$ erg s$^{-1}$, a small fraction of the
  quasar bolometric luminosity. Only about 0.07 per cent of this jet power is
  released as radiation, leaving most available for
  heating the intergalactic medium, and the radiated power ratio is even lower if the
jet contains heavy particles.  \source's core X-ray spectrum is typical
  of other high-redshift quasars, for the majority of which the
  extended X-ray jet properties are unknown, and variability constrains the
emitting region to be less than about 0.7~pc in size.  \source\ also falls into the
  significant group of radio-loud quasars at moderate to high redshift
  for which the nuclear spectrum shows significant curvature implying
  intrinsic X-ray absorption of $N_{\rm H}$ a few $10^{22}$
  cm$^{-2}$ if at the redshift of the source.

\section*{Acknowledgements}

Results are based on observations with \chandra, the \vla\ and \hst.
We are grateful to the \chandra\ X-ray Center for its support of
\chandra\ and relevant analysis software.  The National Radio
Astronomy Observatory is a facility of the National Science Foundation
operated under cooperative agreement by Associated Universities, Inc.
Data from the NASA/ESA Hubble Space Telescope were obtained from the
Space Telescope Science Institute (STScI), which is operated by the
Association of Universities for Research in Astronomy, Inc., under
NASA contract NAS 5-26555.  This research has made use of the
NASA/IPAC Extragalactic Database (NED), which is operated by the Jet
Propulsion Laboratory, California Institute of Technology, under
contract with the National Aeronautics and Space Administration.
DAS and AS acknowledge support from NASA grant GO8-19085X, NASA contract
NAS8-03060 to the CXC, and grant HST-GO-15376.002-A from the STScI. 
We thank the referee for constructive comments.

\section*{Data availability}

The data underlying this work are publicly accessible from
https://cda.harvard.edu/chaser/ (\chandra),
https://archive.nrao.edu/archive/advquery.jsp (\vla), and
https://archive.stsci.edu/hst/search.php (\hst) using the identifiers
given in the article.  Data products arising from the further
processing described, and used for the figures shown, are available
upon reasonable request to the first author.

\appendix
\section{Multiwavelength spectra}
\label{sec:appendix}

%------------ begin figure
\begin{figure*}
\centering
\includegraphics[width=0.32\linewidth]{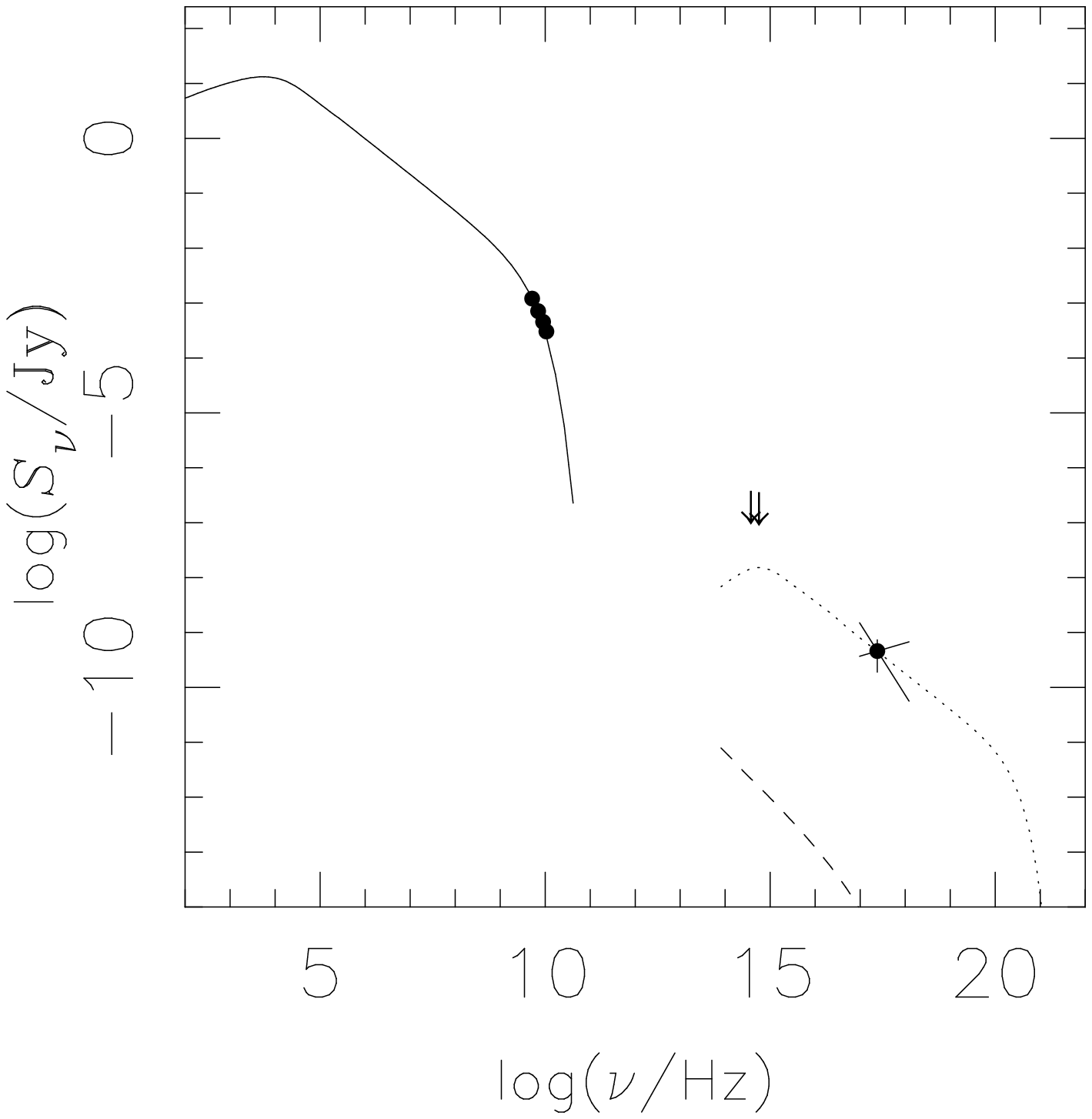}
\includegraphics[width=0.32\linewidth]{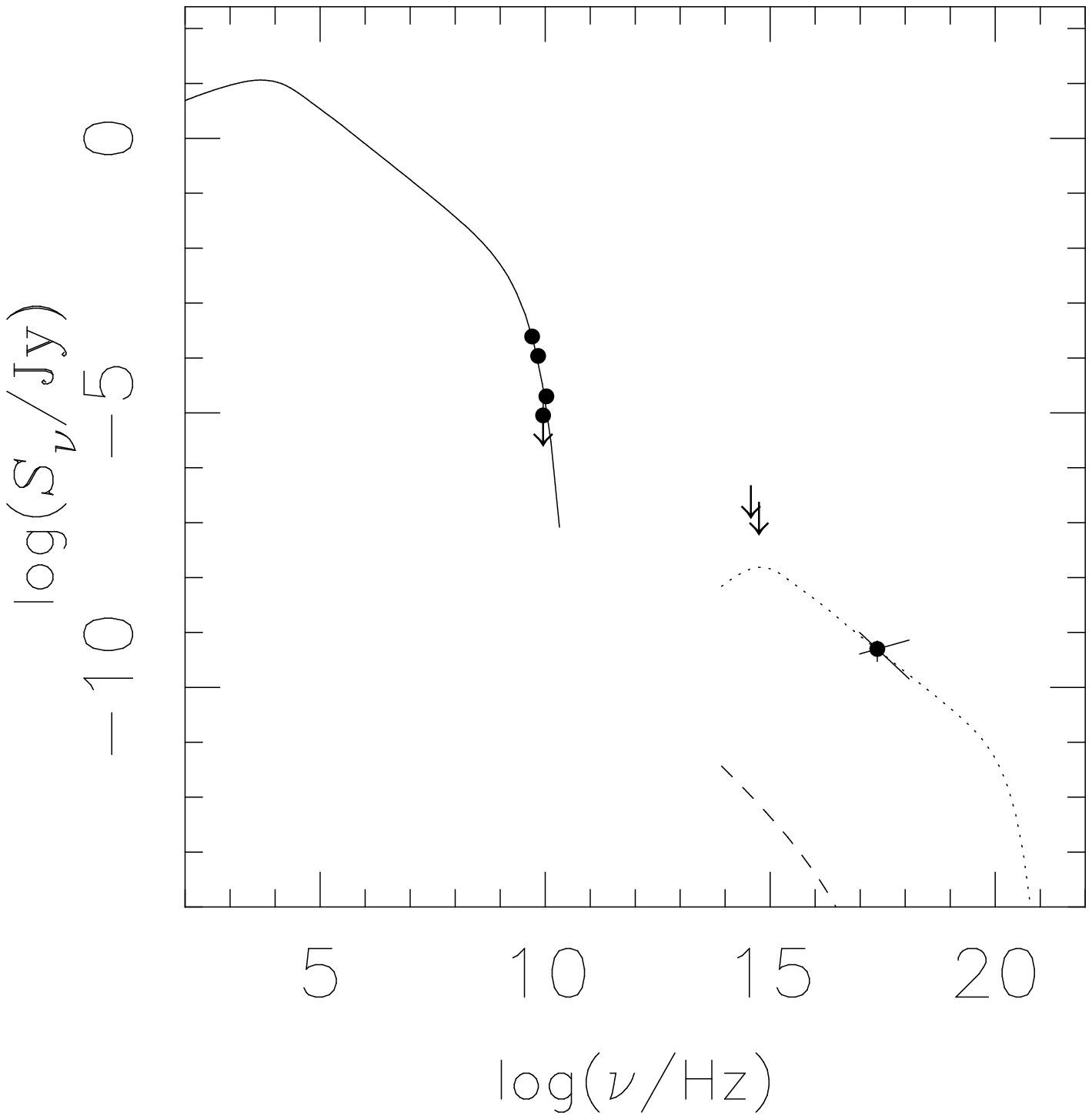}
\includegraphics[width=0.32\linewidth]{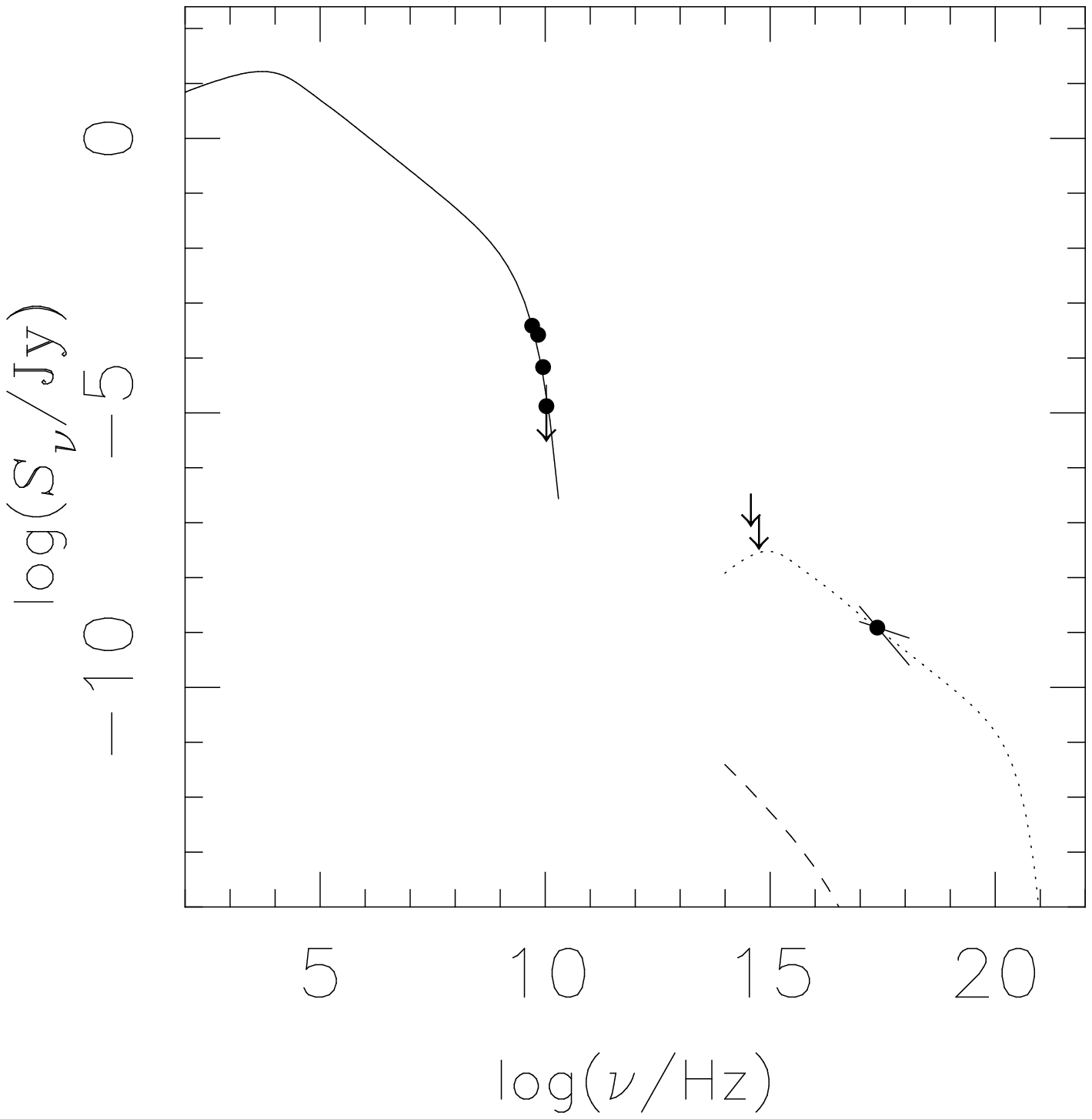}
\caption{Same as Figure~\ref{fig:jetdsed} for jet regions A (left panel), B (centre panel) and C (right panel).}
\label{fig:appendix}
\end{figure*}
%------------ end figure

Multiwavelength spectra similar to Figure~\ref{fig:jetdsed} for the other jet regions are shown in Figure~\ref{fig:appendix}.

\end{document}